\pdfoutput=1

\documentclass[11pt,twoside,a4paper,cmspaper,final,collab]{cms-tdr}
\def\svnVersion{5b9d317-D}\def\svnDate{2020/07/15}\def\cmsCernNoTag{CERN-EP-2020-079}\def\cmsCernDate{\today}\def\cmsMessage{"Published in Physical Review D as \href{http://dx.doi.org/10.1103/PhysRevD.102.032003}{\doi{10.1103/PhysRevD.102.032003}.}"}

\begin{document}\cmsNoteHeader{HIG-18-013}

\hyphenation{had-ron-i-za-tion}
\hyphenation{cal-or-i-me-ter}
\hyphenation{de-vices}
\newcommand{\bb}{\ensuremath{\PQb\PQb}\xspace}
\newcommand{\bbZZ}{\ensuremath{\PQb\PQb\PZ\PZ}\xspace}
\newcommand{\lljj}{\ensuremath{\PQb\PQb\ell\ell\mathrm{jj}}\xspace}
\newcommand{\llnunu}{\ensuremath{\PQb\PQb\ell\ell\nu\nu}\xspace}
\newcommand{\hbb}{\ensuremath{\Ph\to\PQb\PQb}\xspace}
\newcommand{\dyJets}{\ensuremath{\PZ/\gamma^{*}\text{+jets}}\xspace}
\cmsNoteHeader{AN-16-434}
\title{Search for resonant pair production of Higgs bosons in the \texorpdfstring{\bbZZ}{bbZZ} channel in proton-proton collisions at \texorpdfstring{$\sqrt{s}=13$\TeV}{sqrt(s) = 13 TeV}}

\date{\today}
\abstract{A search for the production of a narrow-width resonance decaying into a pair of Higgs bosons decaying into the \bbZZ channel is presented. The analysis is based on data collected with the CMS detector during 2016, in proton-proton collisions at the LHC, corresponding to an integrated luminosity of 35.9\fbinv. The final states considered are the ones where one of the \PZ bosons decays into a pair of muons or electrons, and the other \PZ boson decays either to a pair of quarks or a pair of neutrinos. Upper limits at 95\% confidence level are placed on the production of narrow-width spin-0 or spin-2 particles decaying to a pair of Higgs bosons, in models with and without an extended Higgs sector.  For a resonance mass range between 260 and 1000\GeV, limits on the production cross section times branching fraction of a spin-0 and spin-2 resonance range from 0.1 to 5.0\unit{pb} and 0.1 to 3.6\unit{pb}, respectively.  These results set limits in parameter space in bulk Randall--Sundrum radion, Kaluza--Klein excitation of the graviton, and N2HDM models. For specific choices of parameters the N2HDM can be excluded in a mass range between 360 and 620\GeV for a resonance decaying to two Higgs bosons. This is the first search for Higgs boson resonant pair production in the \bbZZ  channel.}

\hypersetup{%
pdfauthor={CMS Collaboration},%
pdftitle={Search for resonant pair production of Higgs bosons in the bbZZ channel in proton-proton collisions at sqrt(s)=13 TeV},%
pdfsubject={CMS},%
pdfkeywords={CMS, physics, Higgs, b quark, BSM}}

\maketitle

\section{Introduction}
\label{introduction}

The discovery of the Higgs boson ($\Ph$) in 2012~\cite{Chatrchyan:2012xdj,Aad:2012tfa,CMS:2012nga,ATLASscience} has led to a detailed program of studies of the Higgs field couplings to the elementary particles of the standard model (SM) of particle physics: leptons, quarks, and gauge bosons. To fully understand the form of the Higgs field potential, which is a key element in the formulation of the SM, it is important to also study the self-interaction of the Higgs boson. The self-interaction can be investigated through measurements of the production of a pair of Higgs bosons ($\Ph\Ph$). In the SM, $\Ph\Ph$ production is a rare, nonresonant process, with a small production rate~\cite{HiggsXS} that will require the future data sets of the high-luminosity LHC to be observed~\cite{HiggsXS}. Hence, an early observation of $\Ph\Ph$ production, a resonant production in particular, would be a spectacular signature of physics beyond the standard model (BSM). The production of gravitons, radions, or stoponium~\cite{Tang:2012pv,PhysRevD.63.056007,PhysRevD.90.055007}, for example, could lead to $s$-channel $\Ph\Ph$ production via narrow-width resonances. The breadth of the Higgs boson decay channels provides a unique opportunity to test the self-consistency of an $\Ph\Ph$ signal with the SM or models with extended electroweak sectors, such as two-Higgs doublet models (2HDM)~\cite{2HDM1, 2HDM2} or extensions of the minimal supersymmetric standard model~\cite{hH2,hH3,hH5}.

This paper reports a search for resonant $\Pp\Pp\to\mathrm{X}\to\PH\PH$ production in the $\PH\PH\to\bbZZ$ decay channel, where X is a narrow-width resonance of spin-0 or spin-2, and $\PH$ can represent either $\Ph$ or an additional Higgs boson from an extended electroweak sector. The search uses proton-proton ($\Pp\Pp$) collision data at $\sqrt{s}=13\TeV$, recorded with the CMS detector at the LHC in 2016, and corresponding to an integrated luminosity of 35.9\fbinv. It covers a range of resonance masses between 260 and 1000\GeV. The final state consists of two \PQb jets from one Higgs boson decay and two distinct $\PZ$ boson decay signatures from the other $\PH\to\PZ\PZ$ decay: two same-flavor, opposite-sign (OS) leptons from a decay of one of the \PZ bosons, and either two jets of any flavor (the \lljj channel) or significant missing transverse momentum (the \llnunu channel) from the decay of the second \PZ boson to neutrinos. In both cases, the selected charged leptons are either electrons or muons. In the SM, the branching fractions of these signatures represent 0.43 (0.12)\% of the full $\Ph\Ph$ decay through the \bbZZ  intermediate state in the \lljj (\llnunu{}) channel. The challenging aspect of the search in the \lljj channel is the ability to discriminate the signal containing two \PQb jets and two additional jets from multijet background events. For a search in the \llnunu channel, the challenge resides in discriminating the signal against top quark anti-top quark ($\ttbar$) events and instrumental background sources of large missing transverse momentum arising from the mismeasurement of the energies of jets in the final state. The two channels are kept independent by applying orthogonal selections on the missing transverse momentum of the event. Signal yields are calculated for each individual channel and are then combined. Having multiple decay channels with complementary background compositions and sensitivities over a large resonance mass ($m_X$) range makes this combination of the \llnunu and \lljj channels highly efficient for covering the \bbZZ  final state. This is the first search for Higgs boson resonant pair production in the \bbZZ  channel.

Previous searches for resonant $\Ph\Ph$ production have been performed by the CMS and ATLAS Collaborations in the $\bb\bb$~\cite{hh4b,Aaboud:2018knk}, $\bb\tau\tau$~\cite{bbtautau,Aaboud:2018bun}, $\bb\gamma\gamma$~\cite{bbgammagamma}, and $\bb\ell\nu\ell\nu$~\cite{bbWW,Aaboud:2018bun} channels. While coverage of as many $\Ph\Ph$ decay channels as possible remains necessary to understand the exact nature of the Higgs boson self-coupling and the electroweak symmetry breaking mechanism, a \bbZZ  search is particularly interesting in models with extended electroweak sectors, where the phenomenology of additional Higgs bosons can lead to significantly enhanced \bbZZ  production, while suppressing the BSM production of $\bb\bb$, $\bb\tau\tau$, or $\bb\gamma\gamma$ final states.

\section{Benchmark models}
\label{benchmarks}

As in the previous searches, a class of narrow width resonance models arising from the Randall--Sundrum (RS) model~\cite{RandallSundrum} in warped extra dimensions~\cite{radion1,radion2,radion3,radion4} are considered. This scenario introduces one small spatial extra dimension with a nonfactorizable geometry, where the SM particles are not allowed to propagate along that extra dimension, and is referred to in this search as RS1. The resonant particle can be a radion (spin-0) or the first Kaluza--Klein (KK) excitation of a graviton (spin-2). The production cross section of the radion is proportional to $1/\lambda_{\mathrm{R}}^{2}$ where $\lambda_{\mathrm{R}}$ is the interaction scale parameter of the theory. In this analysis, we consider the cases where $\lambda_{\mathrm{R}}=1\TeV$ with $kL = 35$, where $k$ is the constant in the warp factor ($e^{-kL}$) appearing in the space-time metric of the theory and $L$ is the size of the extra dimension. The free parameter of the model for the graviton case is $\tilde{k} = k/\overline{M}_{\mathrm{Pl}}$, where $\overline{M}_{\mathrm{Pl}}$ is the reduced Planck scale, and we consider $\tilde{k}=0.1$ in this analysis~\cite{Oliveira:2014kla}. We further scan the model parameter space in the $\lambda_{\mathrm{R}}$ and $\tilde{k}$ parameters for their respective models. Production at hadron colliders is expected to be dominated by gluon-gluon fusion, and we assume that the radion or graviton is produced exclusively via this process. Due to the small branching fraction of $\Ph\Ph\to\bbZZ$ and the high multiplicities of the final states, the analyses presented in this paper are less sensitive to these models compared to the previous searches. As noted in Section~\ref{introduction}, however, certain models with extended electroweak sectors can produce significantly enhanced \bbZZ  production, while suppressing final states with Higgs boson decays to fermions and scalar bosons.

Such an enhancement can be produced for example in the next-to-minimal 2HDM (N2HDM) extended Higgs sector~\cite{N2HDM,N2HDM1}, where an additional real singlet is introduced in addition to the usual two doublet Higgs bosons of the 2HDM. This analysis is further interpreted in this context. The so-called broken phase is considered, wherein both the Higgs doublets and the singlet acquire vacuum expectation values (vev)~\cite{N2HDM1}. Mixing between these states produces 3 $CP$-even Higgs bosons ${\PH}_1$, ${\PH}_2$, and ${\PH}_3$, with masses that are free parameters of the model. This search considers the nearly mass-degenerate case where the masses of the two bosons ${\PH}_1$ and ${\PH}_2$ are constrained to the experimental measurements of the $\Ph$ mass, which would be indistinguishable from $\Ph$ production with current LHC data sets~\cite{hH1,hH4,hH2}, but may give rise to manifestly non-SM-like rates in the case of $\Ph\Ph$ production. In what is commonly referred to as Higgs boson cascade decays, ${\PH}_3$ can decay to any combination of bosons ${\PH}_1$ and ${\PH}_2$, which then both can have different decay branching fractions compared to the SM Higgs boson. The model spectrum depends on the ratio of the vevs of the two Higgs doublets $\tan\beta$, low values of which enhance ${\PH}_3$ production; the vev of the singlet, which affects the decay branching fractions of ${\PH}_3$ to ${\PH}_1$ and ${\PH}_2$; and three mixing angles, which determine the decay branching fractions of ${\PH}_1$ and ${\PH}_2$~\cite{N2HDM1}. The model spectra described below are determined using \textsc{N2HDECAY}~\cite{N2HDM2}, and are chosen to enhance production of the \bbZZ  final state while respecting current LHC measurements of the SM $\Ph$ branching fractions within their experimental uncertainties~\cite{HiggsXS}. The gluon-gluon fusion production cross sections of ${\PH}_3$ are determined from the BSM Higgs boson predictions of the LHC Higgs Cross Section Working Group~\cite{HiggsXS}. These cross sections assume SM decay branching fractions of the Higgs boson, and changing these branching fractions affects the production cross section. The cross sections are corrected at leading order (LO) by the ratio of the relative partial width of ${\PH}_3$ in the decay to two gluons compared to the BSM Higgs boson prediction. Enhanced (reduced) coupling of ${\PH}_3$ to gluons will enhance (reduce) the production cross section of ${\PH}_3$. The mass of the Higgs bosons ${\PH}_1$ and ${\PH}_2$ are set to 125\GeV, and the mass of ${\PH}_3$ is generated in the range $260\leq m_\mathrm{X}\leq 1000\GeV$. Two benchmark points are chosen for this analysis, corresponding to $\tan\beta=0.5$ and 2.0. In both cases, the scalar vev is set to 45\GeV, and the mixing angles $\alpha_1$, $\alpha_2$, $\alpha_3$ are set to 0.76, 0.48, and 1.00, respectively. For $\tan\beta=0.5$, this results in branching fractions of ${\PH}_3$ to ${\PH}_1{\PH}_1$, ${\PH}_1{\PH}_2$, and ${\PH}_2{\PH}_2$ around 0.02, 0.29, and 0.64 respectively, branching fractions of ${\PH}_1\to\PQb\PQb$ (${\PH}_1\to\PZ\PZ$) of 0.70 (0.01), and branching fractions of ${\PH}_2\to\PQb\PQb$ (${\PH}_2\to\PZ\PZ$) of 0.42 (0.05). This represents a 33\% increase in the branching fraction to \bbZZ  compared to SM $\Ph\Ph$ decays. The correction factor based on the relative partial width of ${\PH}_3$ to two gluons is around 3.0. For  $\tan\beta=2.0$, this results in branching fractions of ${\PH}_3$ to ${\PH}_1{\PH}_1$, ${\PH}_1{\PH}_2$, and ${\PH}_2{\PH}_2$ around 0.07, 0.22, and 0.67 respectively, branching fractions of ${\PH}_1\to\PQb\PQb$ (${\PH}_1\to\PZ\PZ$) of 0.53 (0.03), and branching fractions of ${\PH}_2\to\PQb\PQb$ (${\PH}_2\to\PZ\PZ$) of 0.58 (0.03). This represents a 5\% increase in the branching fraction to \bbZZ  compared to SM $\Ph\Ph$ decays. The correction factor based on the relative partial width of ${\PH}_3$ to two gluons is around 0.7. These corrections and branching fractions produce significant differences in the production rates of the \bbZZ  system compared to $\Ph\Ph$ production both in the SM and through resonant production of radions or gravitons.

\section{The CMS detector}
\label{cms}

The central feature of the CMS apparatus is a superconducting solenoid of 6\unit{m} internal diameter, providing a magnetic field of 3.8\unit{T}. Within the solenoid volume are a silicon pixel and strip tracker, a lead tungstate crystal electromagnetic calorimeter (ECAL), and a brass and scintillator hadron calorimeter (HCAL), each composed of a barrel and two endcap sections. Forward calorimeters extend the pseudorapidity coverage provided by the barrel and endcap detectors, where pseudorapidity is defined as $\eta = -\ln[\tan(\theta /2)]$, and $\theta$ is the polar angle. Muons are measured in gas-ionization detectors embedded in the steel flux-return yoke outside the solenoid. CMS uses a two-level trigger system~\cite{Khachatryan:2016bia}. The first level of the CMS trigger system, composed of custom hardware processors, uses information from the calorimeters and muon detectors to select the most interesting events. The high-level trigger (HLT) processor farm further decreases the event rate from around 100\unit{kHz} to a rate of around 1\unit{kHz}, before data storage. A more detailed description of the CMS detector, together with a definition of the coordinate system used and the relevant kinematic variables, can be found in Ref.~\cite{Chatrchyan:2008aa}. 

\section{Event simulation}
\label{samples}

The signal samples of RS1 spin-0 radion and RS1 KK spin-2 graviton narrow resonances decaying to a pair of Higgs bosons ($\mathrm{X}\to\Ph\Ph$) are generated at LO using {\MGvATNLO}. The $\Ph$ mass is set to 125\GeV, and the $X$ resonance mass $m_X$ is generated in the range of 260--1000\GeV. In the \llnunu channel the final state can be produced via either the \bbZZ  or $\PQb\PQb\PWp\PWm$ intermediate states.

The main background processes to production of a pair of Higgs bosons in the $\bbZZ\to$ \lljj  or \llnunu final states are \dyJets and $\ttbar$ processes. Less significant backgrounds arise from single top quark, $\PW$+jets, diboson+jets, SM Higgs boson production, and quantum chromodynamics (QCD) multijet production. Signal and background processes are modeled with simulations, with the exception of the QCD multijet background that is estimated using data control regions. 

In the analysis using the \lljj channel, the \dyJets and $\PW$+jets processes are generated with {\MGvATNLO}2.4.2~\cite{Alwall:2014hca} at next-to-leading order (NLO). In this case, the generator uses the \textsc{FxFx} jet merging scheme~\cite{Frederix:2012ps}. The analysis of the \llnunu channel uses samples of \dyJets generated with {\MGvATNLO} at LO, with the MLM matching scheme~\cite{Alwall:2007fs}, and reweighted to account for higher order QCD and electroweak effects~\cite{DY_QCDnEWK}.

{\tolerance=800 The \ttbar process is generated at NLO with \POWHEG2.0~\cite{Nason:2004rx,Frixione:2007vw,Alioli:2010xd,Alioli:2009je,Re:2010bp,Alioli:2008tz}. Single top processes and SM Higgs boson production processes are simulated at NLO either with \POWHEG or {\MGvATNLO}, depending on the particular channel. The diboson processes ($\PW\PW$+jets, $\PW\PZ$+jets, $\PZ\PZ$+jets) are simulated at NLO with {\MGvATNLO}.\par}

The simulated samples are normalized to their best-known highest-order-QCD cross sections, either evaluated at NLO with \MCFM~\cite{Campbell:2010ff} (diboson+jets) or at next-to-next-to-leading order with \FEWZ 3.1~\cite{Li:2012wna} (single top quark, $\PW$+jets, SM Higgs boson), with the exception of $\ttbar$ and \dyJets processes, which are normalized using data.

The simulated samples are interfaced with \PYTHIA8.212~\cite{pythia8p2} for parton showering and hadronization. The \PYTHIA generator uses the CUETP8M1 underlying event tune~\cite{Khachatryan:2015pea}. The \textsc{nnpdf3.0} NLO and LO parton distribution functions (PDFs)~\cite{Ball:2014uwa} are used for the various processes, with the precision matching that in the matrix element calculations.

For all the simulated samples used in this analysis, a simulation of CMS detector response based on \GEANTfour~\cite{GEANT4} is applied. The presence of additional interactions in the same bunch crossing (pileup, or PU), both in-time and out-of-time with respect to the primary interaction, is simulated and corrected to agree with a multiplicity corresponding to the distribution measured in data.

\section{Event reconstruction and background estimation}
\label{selection}

\subsection{Event reconstruction}
\label{reconstruction}
Events are selected using triggers that require two muons with transverse momentum $\pt >17$ (8)\GeV or two electrons with $\pt >23$ (12)\GeV for the leading (sub-leading) lepton. 

The particle-flow (PF) algorithm~\cite{PFlast}, which combines information from various elements of the CMS detector, is used to reconstruct and identify final-state particles, such as photons, electrons, muons, and charged and neutral hadrons, as individual PF objects. Combinations of PF objects are then used to reconstruct higher-level objects such as jets and missing transverse momentum.

Jets are reconstructed from the PF objects, using the anti-\kt~\cite{antikt,fastjet} algorithm with a distance parameter of $R = 0.4$. In order to reduce instrumental backgrounds and the contamination from PU, selected jets are required to satisfy loose identification criteria~\cite{CMS:2017wyc} based on the multiplicities and energy fractions carried by charged and neutral hadrons. The energy of reconstructed jets is calibrated using $\pt $- and $\eta$-dependent corrections to account for nonuniformity and nonlinearity effects of the ECAL and HCAL energy response to neutral hadrons, for the presence of extra particles from PU, for the thresholds used in jet constituent selection, reconstruction inefficiencies, and possible biases introduced by the clustering algorithm. These jet energy corrections are extracted from the measurement of the momentum balance in dijet, $\text{photon}+\text{jet}$, \dyJets, and multijet events~\cite{Khachatryan:2016kdb}. A residual $\eta$- and $\pt $-dependent calibration is applied to correct for the small differences between data and simulated jets. The jets that are candidates to be from the decay of one of the Higgs bosons and of one of the \PZ bosons are required to have $\pt >20\GeV$. Furthermore, jets are required to have a spatial separation of $\Delta R > 0.3$ from lepton candidates. 

Jets originating from \PQb quarks are identified with the combined multivariate analysis (cMVA) algorithm~\cite{Sirunyan:2017ezt}.  A jet is tagged as a \PQb jet if the cMVA discriminant is above a certain threshold, chosen such that the misidentification rate  is about 1\% for light-flavor quark and gluon jets, and about 13\% for charm quark jets. The \PQb jet tagging efficiency for this working point is about 66\%.

The missing transverse momentum vector \ptvecmiss is computed as the negative vector sum of the transverse momenta of all the PF candidates in an event, and its magnitude is denoted as \ptmiss~\cite{Sirunyan:2019kia}. The \ptvecmiss is modified to account for corrections to the energy scale of the reconstructed jets in the event.

The candidate vertex with the largest value of summed physics-object $\pt^2$ is taken to be the primary $\Pp\Pp$ interaction vertex. The physics objects are the jets, clustered using the jet finding algorithm~\cite{antikt,fastjet} with the tracks assigned to candidate vertices as inputs, and the associated missing transverse momentum, taken as the negative vector sum of the \pt  of those jets.

Muons are reconstructed as tracks in the muon system that are matched to the tracks reconstructed in the inner silicon tracking system~\cite{MuId}. The leading muon is required to have $\pt  > 20\GeV$, while the subleading muon must have $\pt  > 15$ (10)\GeV in the \llnunu (\lljj{}) channel. Muons are required to be reconstructed in the HLT fiducial volume, \ie{}, with $\abs{\eta} < 2.4$, to ensure that the offline selection is at least as restrictive as the HLT requirements. The selected muons are required to satisfy a set of identification requirements based on the number of spatial measurements in the silicon tracker and in the muon system and the fit quality of the combined muon track~\cite{Khachatryan:2015pea}, and are required to be consistent with originating from the primary vertex.

Electrons are reconstructed by matching tracks in the silicon tracker to the clusters of energy deposited in the ECAL~\cite{EMId}. The leading (subleading) electron is required to have $\pt  > 25$ (15)\GeV and $\abs{\eta} < 2.5$ to be within the geometrical acceptance, excluding candidates in the range $1.4442 < \abs{\eta} < 1.5660$, which is the transition region between the ECAL barrel and endcaps, because the reconstruction of an electron in this region is poor compared to other regions. Electrons are required to pass an identification requirement based on an MVA~\cite{TMVA} technique that combines information from various observables related to the shower shape in the ECAL and the quality of the matching between the tracks and the associated ECAL clusters~\cite{EMId}. They are further required to be consistent with originating from the primary vertex. Candidates that are identified as originating from photon conversions in the material of the detector are removed.

Both muons and electrons have a requirement that the lepton relative isolation, defined in Eq.(\ref{isoeq}), be less than 0.25 (0.15) and 0.15 (0.06), respectively, for the \lljj (\llnunu{}) channel. In Eq.(\ref{isoeq}), the sums run over charged hadrons originating from the primary vertex of the event, neutral hadrons, and photons inside a cone of radius $\Delta R = \sqrt{\smash[b]{(\Delta\phi )^2 + (\Delta\eta) ^2}} < 0.4$ (0.3) around the direction of the muon (electron), where $\phi$ is the azimuthal angle in radians.
\ifthenelse{\boolean{cms@external}}{
       \begin{multline}
              \label{isoeq}
              I_\text{iso} = \frac{1}{\pt^{\ell}}\Biggl[\sum^{\text{charged}}\pt \\+ \max\Bigl(0, \sum^{\text{neutral}}\pt + \sum^{\text{photons}} \pt - \text{Corr}_{\text{PU}}\Bigr)\Biggr]
       \end{multline}
}{
\begin{equation}
\label{isoeq}
I_\text{iso} = \frac{1}{\pt^{\ell}}\left[\sum^{\text{charged}}\pt + \max\left(0, \sum^{\text{neutral}}\pt + \sum^{\text{photons}} \pt - \text{Corr}_{\text{PU}}\right)\right]
\end{equation}
}
{\tolerance=800 The isolation includes a correction for pileup effects, $\text{Corr}_{\text{PU}}$. For electrons, $\text{Corr}_{\text{PU}}=\rho A_{\text{eff}}$, where $\rho$ is the average transverse momentum flow density, calculated using the jet area method~\cite{PUcorr}, and $A_{\text{eff}}$ is the geometric area of the isolation cone times an $\eta$-dependent correction factor that accounts for residual pileup effects. For muons, $\text{Corr}_{\text{PU}}=0.5\sum^{\text{PU}}\pt $, where the sum runs over charged particles not associated with the primary vertex and the factor 0.5 corresponds to an approximate average ratio of neutral to charged particles in the isolation cone~\cite{CMS:2010eua}.\par}

Simulated background and signal events are corrected with scale factors for differences observed between data and simulation, in trigger efficiencies, in lepton $\pt $- and $\eta$-dependent identification and isolation efficiencies, and in \PQb\ tagging efficiencies. 

\subsection{\texorpdfstring{Event selection in the \lljj channel}{Event selection in the bblljj channel}}
\label{lljjselection}

After selection of the candidate physics objects, an initial event selection is performed by requiring at least two same-flavor leptons (muons or electrons) in each event. The two leptons are required to be oppositely charged. The invariant mass of the two leptons, $m_{\ell\ell}$, is required to be larger than 15\GeV. Four of the jets in an event are designated as the $\Ph$ and \PZ boson decay products. These jets are required to have $\pt >20\GeV$ and at least one of those must be \PQb tagged with a minimum requirement on the \PQb tagging discriminant, that is looser than the requirement in the final selection. We refer to this selection as the preselection.

Since the signal contains two \PQb jets from the decay of a Higgs boson, and two jets of any flavor from the decay of a \PZ boson, it is important to carefully categorize the jets in the event. Starting from a collection of jets identified as described above, the information from the \PQb tagging discriminant, as well as the kinematic properties of the jets, are taken into account when assigning jets as each particle's decay products. 

The following selection is applied to identify the \PQb jets originating from the decay of the Higgs boson. The two jets with the highest \PQb tagging scores above a certain threshold are assigned to the decay of the Higgs boson. If only one jet is found that meets the minimum \PQb tagging score value, a second jet that leads to an invariant mass closest to 125\GeV~is selected. If no jets with \PQb tagging scores above threshold are found, the two jets whose invariant mass is closest to 125\GeV~are chosen. 

After jets are assigned to the decay of \hbb, from the remaining jets the two jets with four-object invariant mass $M(\ell\ell\text{jj})$ closest to 125\GeV are assigned to the decay of the \PZ boson.

After preselection, additional requirements are imposed. At least one of the four jets assigned as the decay products of the $\Ph$ or \PZ boson must satisfy the \PQb tagging requirement, to increase the signal-to-background ratio. To impose orthogonality with the \llnunu decay channel, upper limits on the $\ptmiss$ are imposed as follows: $\ptmiss < 40,\ 75,\ \mathrm{and}\ 100\GeV$ for the $m_X$ of 260--350, 350--650, and ${\geq}$650\GeV, respectively. We refer to this selection as the final selection in the \lljj channel.

After the final selection, twenty-two variables that exploit the differences in kinematic and angular distributions between the signal and background processes are combined into a boosted decision tree (BDT) discriminant~\cite{bdt}. In the $m_X$ range of 260--300\GeV, the most important variables are $m_{\ell\ell}$, the separation between the leading lepton and leading \PQb tagged jet $\Delta R_{\ell1\PQb1}$, and the invariant mass of the pair of \PQb tagged jets $m^{\Ph}_{\PQb\PQb}$. In the $m_X$ range of 350--550\GeV, $m^{\Ph}_{\PQb\PQb}$ is the most important variable, while $m_{\ell\ell}$ becomes less important, and the separation between the pair of leptons $\Delta R_{\ell\ell}$ gradually becomes more important when the $m_X$ increases. For the $m_X$ higher than 550\GeV, $\Delta R_{\ell\ell}$ becomes the most important variable followed by $m^{\Ph}_{\PQb\PQb}$ and the separation between the pair of \PQb tagged jets $\Delta R^{\Ph}_{\PQb\PQb}$. The BDTs are configured to use stochastic gradient boosting with the binomial log-likelihood loss function. The software package TMVA~\cite{TMVA} is used for BDT implementation, training, and application.

The BDT is trained using all background processes described in Section~\ref{samples}, excluding the multijet background. In each lepton channel and for each spin hypothesis, one BDT is trained for each simulated signal $m_X$. In the training, signal events include samples from the two neighboring mass points, in addition to the targeted mass point. In total, 48 BDTs are trained. These BDT distributions for data and expected backgrounds are used as the final discriminating variable in the analysis.

\subsection{\texorpdfstring{Background estimation in the \lljj channel}{Background estimation in the bblljj channel}}
\label{lljjbackground}

The main processes that can mimic the signature of the signal in the \lljj channel are \dyJets and $\ttbar$, with smaller contributions from QCD multijets, diboson+jets, $\PW$+jets, and SM Higgs boson production. 

The contribution from the principal background, \dyJets, is estimated with simulated events normalized to the data at the preselection level in the $\PZ$ boson-enriched control region $80 < m_{\ell\ell} < 100\GeV$. The contribution from $\ttbar$ is estimated in a similar manner, with the $\ttbar$-enriched control region defined by $m_{\ell\ell} > 100\GeV$, and $\ptmiss>100\GeV$. The data-to-simulation normalization factors derived from the two control regions are $R_{\PZ} = 1.14 \pm 0.01\stat$ and $R_{\ttbar} = 0.91 \pm 0.01\stat$ in the muon channel and $R_{\PZ} = 1.24 \pm 0.01\stat$ and $R_{\ttbar} = 0.97 \pm 0.02\stat$ in the electron channel. These normalization factors are found to be consistent between lepton flavors when applying lepton-specific systematic variations.

The contribution from QCD multijet processes is determined from data with a method that exploits the fact that neither signal events nor events from other backgrounds produce final states with same-sign leptons at any significant level. Data events with same-sign isolated leptons are used to model the shape of the multijet background, after all non-QCD sources of background contributing to this selection are subtracted using simulation. The yield in this region is normalized with the ratio of the number of events with nonisolated OS leptons to the number of events with nonisolated same-sign leptons. Here, nonisolated leptons are those muons (electrons) that fail the relative isolation requirements described in Section~\ref{reconstruction}. All non-QCD sources of background, estimated with simulated events, are subtracted from the numerator and the denominator before computing the ratio. 

The contributions from diboson+jets, $\PW$+jets, and SM Higgs boson production are estimated from simulation.

\subsection{\texorpdfstring{Event selection in the \llnunu channel}{Event selection in the bbllnunu channel}}
\label{llnunuselection}

Candidate events in the  \llnunu channel are reconstructed from the physics objects, as described above. The two leptons (muons or electrons) are required to have OS, and the invariant mass of the two leptons, $m_{\ell\ell}$, is required to exceed 76\GeV. One of the Higgs bosons is formed from the pair of \PQb jets with the highest output value of the \PQb tagging discriminant, and the second Higgs boson is reconstructed as a combination of the two charged leptons and the \ptvecmiss, representing the visible and invisible decays products, respectively, of the pair of \PZ bosons. The requirement on $m_{\ell\ell}$ reduces the contribution from resonant $\mathrm{X}\to\Ph\Ph$ production in the $\PQb\PQb\PW\PW$ final state, and makes this measurement orthogonal to a previous $\PQb\PQb\PW\PW$ search~\cite{bbWW}, where only events with $m_{\ell\ell}$ below 76\GeV were considered.

 For the Higgs boson decaying to a pair of \PZ bosons, the two neutrinos are not reconstructed in the detector, and a pseudo invariant mass of the Higgs boson is used to approximate the incomplete momentum four-vector of the $\PH$. The pseudo invariant mass is formed from the momenta of the two charged leptons coming from one of the \PZ bosons and the four-vector $(\ptmiss, \ptvecmiss)$ approximating that of the two-neutrino system coming from the other of the \PZ bosons, where the $z$ component of \ptvecmiss is zero. While the true invariant mass of the pair of neutrinos is not zero but is equal to the invariant mass of the parent \PZ boson, that boson is off the mass shell and has relatively low mass.

In order to suppress the backgrounds from the \dyJets and QCD multijet processes as well as from the SM Higgs boson production via the $\PZ\Ph$ process, a requirement is imposed on the minimum \ptmiss, which is 40 (75)\GeV for the $m_X$ of 260--300 (350--600)\GeV, and 100\GeV for higher $m_X$.

Three regions, a signal region (SR) and two control regions (CRs), are further defined using $m_{\ell\ell}$ and the invariant mass $m^{\Ph}_{\PQb\PQb}$ of the two \PQb jets. The SR is defined by the requirements $76 < m_{\ell\ell} <106\GeV$ and $90< m^{\Ph}_{\PQb\PQb} <150\GeV$. A first CR, dominated by \ttbar events, is defined by $m_{\ell\ell} >106\GeV$ and $90< m^{\Ph}_{\PQb\PQb} <150\GeV$. A second CR, enriched in \dyJets events, is defined by requiring $20< m^{\Ph}_{\PQb\PQb} <90\GeV$ or $m^{\Ph}_{\PQb\PQb} >150\GeV$, and $76 < m_{\ell\ell} <106\GeV$. The two CRs and the SR are used to estimate the backgrounds in the SR via a simultaneous fit.

To further differentiate signal from backgrounds in the SR, a BDT discriminant is trained using all simulated signal and background processes described in Section~\ref{samples}. Of the nine input distributions to the BDT, the most important variables in the low-mass range are the separation between the pair of $\PQb$ tagged jets $\Delta R^{\Ph}_{\PQb\PQb}$, \ptmiss, and $m^{\Ph}_{\PQb\PQb}$. In the high-mass region, $m^{\Ph}_{\PQb\PQb}$ and $\Delta R^{\Ph}_{\PQb\PQb}$ are also the most significant, together with the separation between the pair of charged leptons $\Delta R_{\ell\ell}$, which becomes more important as the resonance mass increases. Two BDTs are trained for each lepton channel and each resonance spin hypothesis, one for $m_X$ in the range of 250--450\GeV, and another one for the $m_X$ above 450\GeV. A minimum BDT value is required for candidates in the SR, optimized for each narrow $m_X$ hypothesis to yield the best 95\% confidence level (\CL) expected upper limit on resonant production. The BDTs are configured with the same classification and loss function parameters described in Section~\ref{lljjselection}.

Finally, a quantity closely correlated with the energy-momentum four-vector of the $\Ph\Ph$ system is constructed as the vector sum of the of the two leptons, two $\PQb$ jets, and the four-vector formed as $(\ptmiss, \ptvecmiss)$ for the neutrinos, as described above. Subsequently, the pseudo transverse mass of the $\Ph\Ph$ system is defined as $\widetilde{M}_{\mathrm{T}}(\Ph\Ph) = \sqrt{\smash[b]{E^2 - p_{z}^2}}$, where $E$ and $p_{z}$ are the energy and the $z$ component of the combined four-vector.

The $\widetilde{M}_{\mathrm{T}}(\Ph\Ph)$ distributions for data and expected backgrounds, in the combined signal and CRs, will be used as the final discriminating variable in the analysis.

After the event selection in this channel is applied, the signal $\Ph\Ph$ events in the SR come predominantly from the decays with the \bbZZ  intermediate state (80\%) with a smaller contribution from the $\PQb\PQb\PWp\PWm$ intermediate state (20\%). Both intermediate states are used to calculate the limit on $\Pp\Pp\to\mathrm{X}\to\Ph\Ph$ in the \llnunu channel.

\subsection{\texorpdfstring{Background estimation in the \llnunu channel}{Background estimation in the bbllnunu channel}}
\label{llnunubackground}

The dominant sources of background in the \llnunu channel are \ttbar and \dyJets production. Several other background processes contribute, including single top quark and diboson production, and SM Higgs boson production in association with a \PZ boson. While these are typically minor backgrounds, their contribution can vary over the $m_X$ range. The QCD multijet background is negligible across the full mass range because of the stringent  selection on $m_{\ell\ell}$.

The event yields in the signal and two CRs, which are dominated by \ttbar and \dyJets events, are determined from data.  The corresponding normalizations of the simulated $\widetilde{M}_{\mathrm{T}}(\Ph\Ph)$ distributions are free parameters in the simultaneous fit of all three regions. The remaining backgrounds are estimated from simulation and normalized according to their theoretical cross sections. 

\section{Systematic uncertainties}
\label{systematics}

The dominant source of systematic uncertainty in this analysis is the jet energy scale (JES) uncertainty, which is of the order of a few percent and is estimated as a function of jet \pt and $\eta$~\cite{Khachatryan:2016kdb}. The $\eta$-dependent jet energy resolution (JER) correction factors are varied by $\pm$1 standard deviation in order to estimate the effect of the uncertainty. Uncertainties in the JES are propagated to the calculation of $\ptmiss$. A residual $\ptmiss$ uncertainty of 3\% is applied in the \llnunu channel to take into account the effect, at low $\ptmiss$, of the unclustered energy from neutral hadrons and photons that do not belong to any jet, and from jets with $\pt <10\GeV$.

An uncertainty of 2\% per muon in the muon reconstruction, identification, and isolation requirements, as well as a 1\% per muon uncertainty in the muon HLT efficiency are assigned~\cite{MuId}. A per-muon uncertainty due to measured differences of tracking efficiency in data and simulation is estimated to be 0.5\% for muon $\pt <300\GeV$ and 1.0\% for muon $\pt >300\GeV$~\cite{wprime}. Per-electron uncertainties in the efficiency for electron trigger, identification and isolation requirements, estimated by varying the scale factors within their uncertainties, are applied. The uncertainties in the efficiency scale factors are generally $<$2\% for trigger and $<$3\% for identification and isolation~\cite{EMId}. The effect of the variations on the yield of the total background is $<$1\%. Uncertainties in the data-to-simulation correction factors of the \PQb tagging and of light flavor mis-tagging efficiencies are included. 

Normalization and shape uncertainties are assigned to the modeling of the backgrounds. An uncertainty in the shape of the signal and background models is determined by varying the factorization and the renormalization scales between their nominal values and 0.5 to 2.0 times the nominal values in the simulated signal and background samples. The variations where one scale increases and the other decreases are not considered. Each of the remaining variations of the renormalization and the factorization scales are considered, and the maximum variation among all the samples with respect to the nominal sample used in the analysis is taken as the systematic uncertainty, which is found to be 5--7\% depending on the process. An uncertainty in the signal acceptance and background acceptance and cross section due to PDF uncertainties and to the value chosen for the strong coupling constant is estimated by varying the NNPDF set of eigenvectors within their uncertainties, following the PDF4LHC prescription~\cite{PDF4LHC}. Statistical uncertainties in the simulated samples for \dyJets and $\ttbar$ background estimates result in uncertainties on the data-derived normalization factors in the \lljj channel.

An uncertainty of 2.5\% is assigned to the determination of the integrated luminosity~\cite{lumi2016}. The uncertainty in the PU condition and modeling is assessed by varying the inelastic $\Pp\Pp$ cross section from its central value by $\pm$4.6\%~\cite{pu2016}.

All the uncertainties discussed are applied to all background and signal simulated samples. The sensitivity of the presented search is limited by the statistical uncertainties.

\section{Results}
\label{results}

Results are obtained by performing a binned maximum likelihood fit of the BDT distributions for the \lljj channel, and of the $\Ph\Ph$ pseudo transverse mass simultaneously in the SR and two CRs for the \llnunu channel.

\begin{figure*}[htbp]
       \centering
       {\includegraphics[width=.49\textwidth]{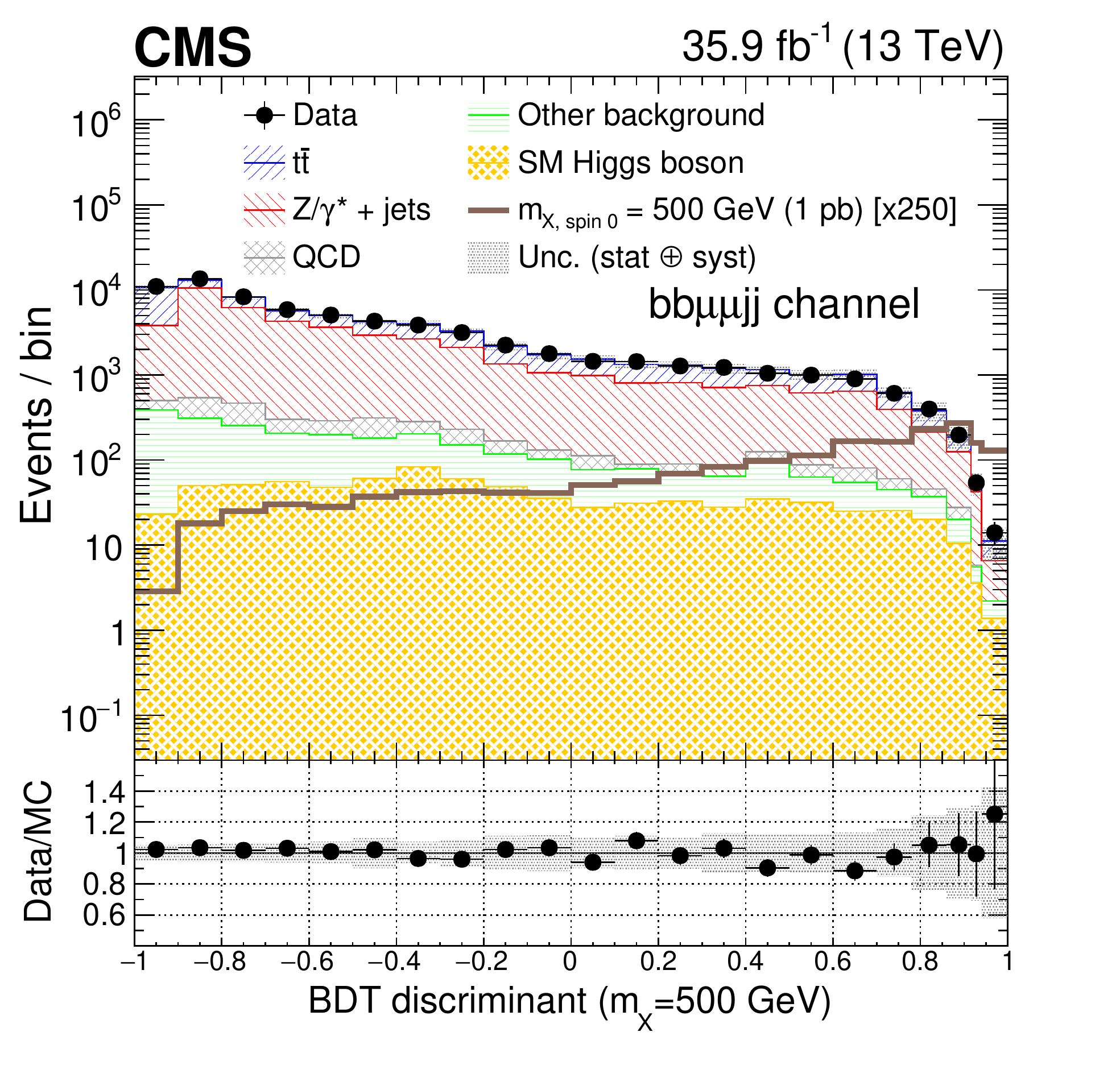}}
       {\includegraphics[width=.49\textwidth]{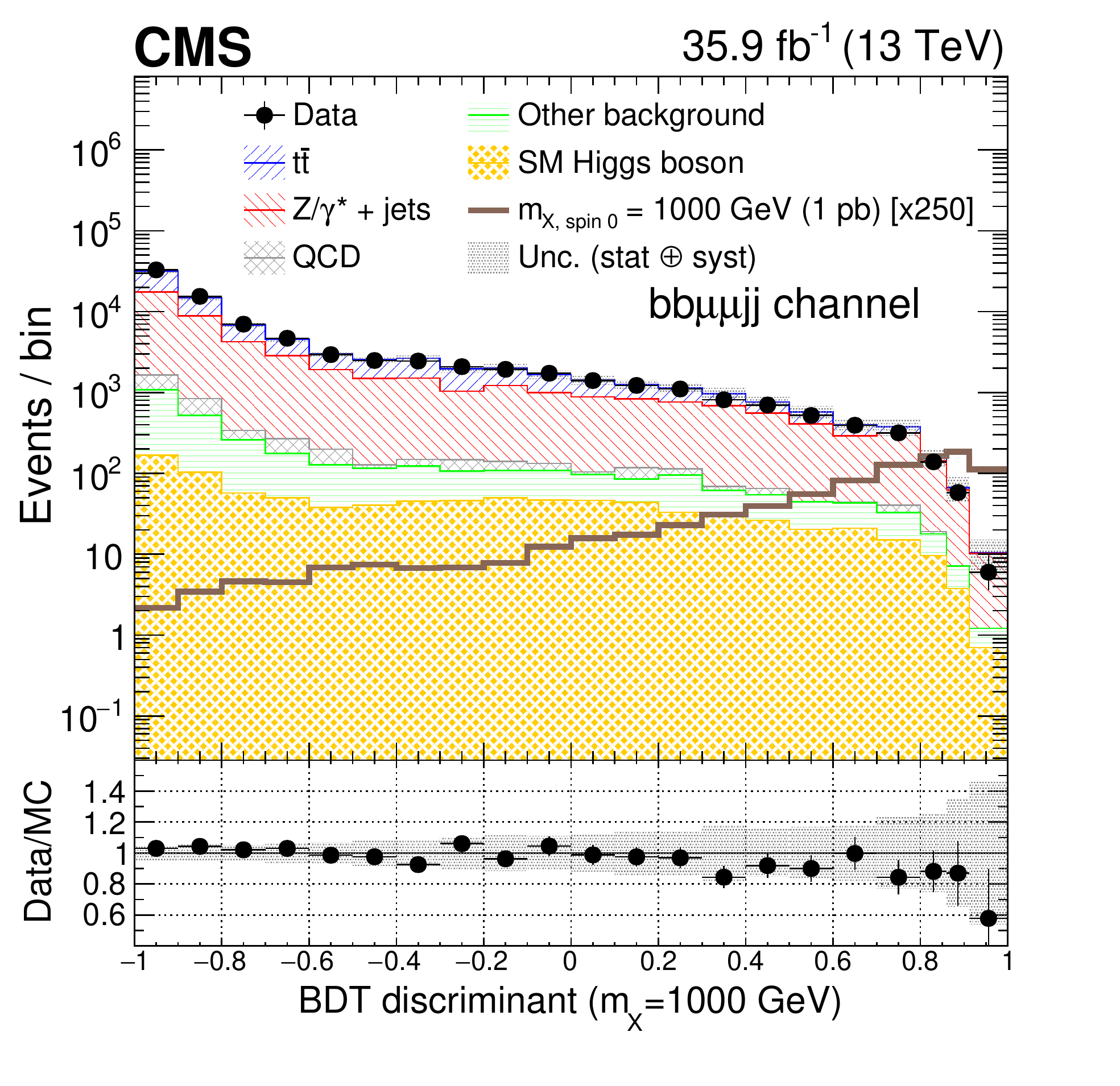}}  
       {\includegraphics[width=.49\textwidth]{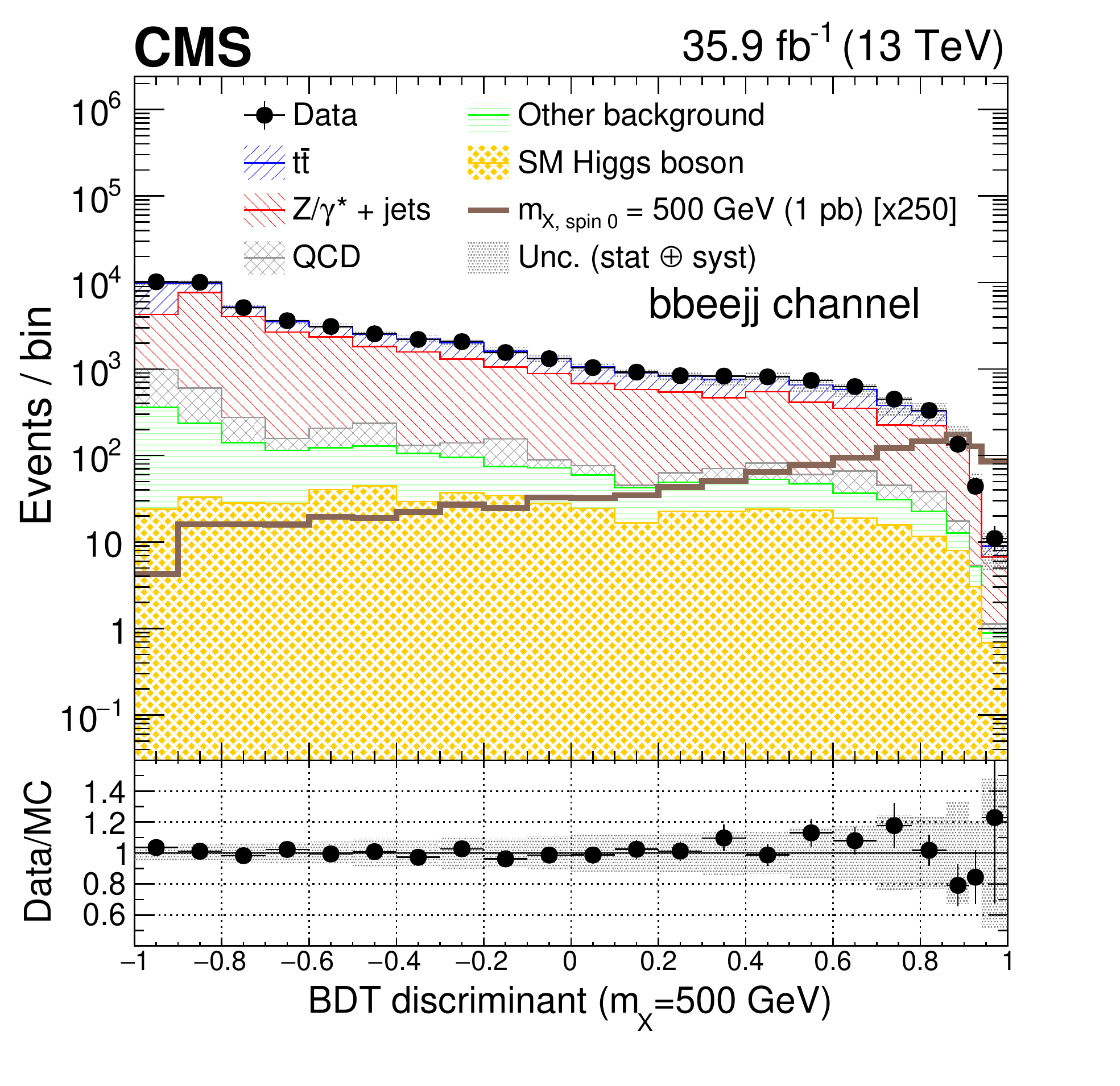}}
       {\includegraphics[width=.49\textwidth]{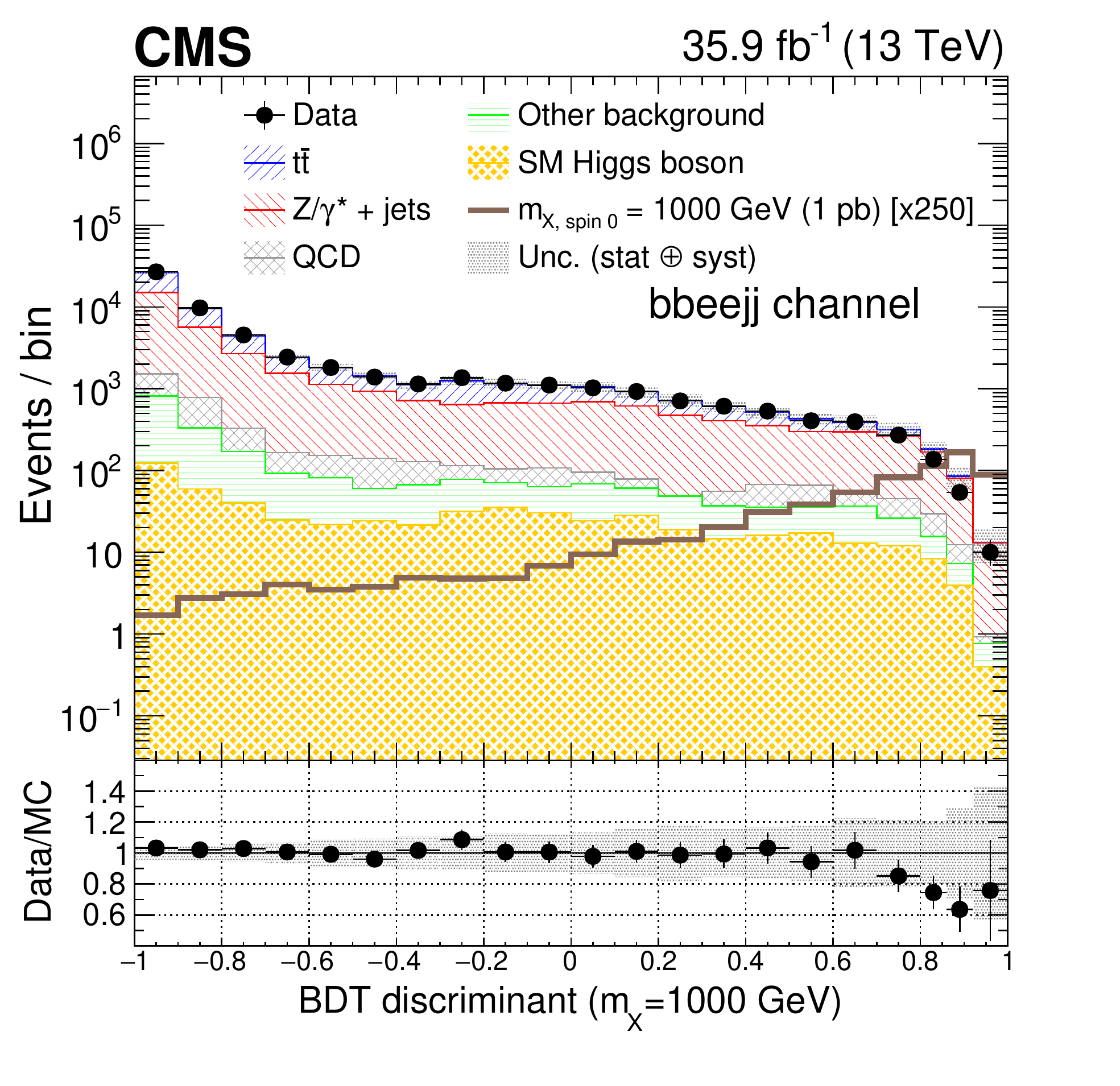}}       
       \caption{Comparison of the BDT discriminant for $m_\mathrm{X} =500$ and 1000\GeV after the final selection in the muon (upper row) and electron (lower row) final states of the \lljj channel. The signals of an RS1 radion with mass of 500 (left) and 1000\GeV (right) are normalized to a cross section of 1\unit{pb} for the $\Pp\Pp\to\mathrm{X}\to\Ph\Ph$ process. The shaded area represents the combined statistical and systematic uncertainties in the background estimate.}
	  \label{figapp:bdt}
\end{figure*}

The data and background predictions at final selection level in the \lljj channel are shown in Fig.~\ref{figapp:bdt}, for the distributions of the BDT discriminant for signal masses of 500 and 1000\GeV, in the muon and electron final states. Studies performed on all 48 BDT discriminants show stability of the trainings with no evidence of bias or overtraining.

\begin{figure*}[tbp]
  \centering
    \includegraphics[width=0.32\textwidth]{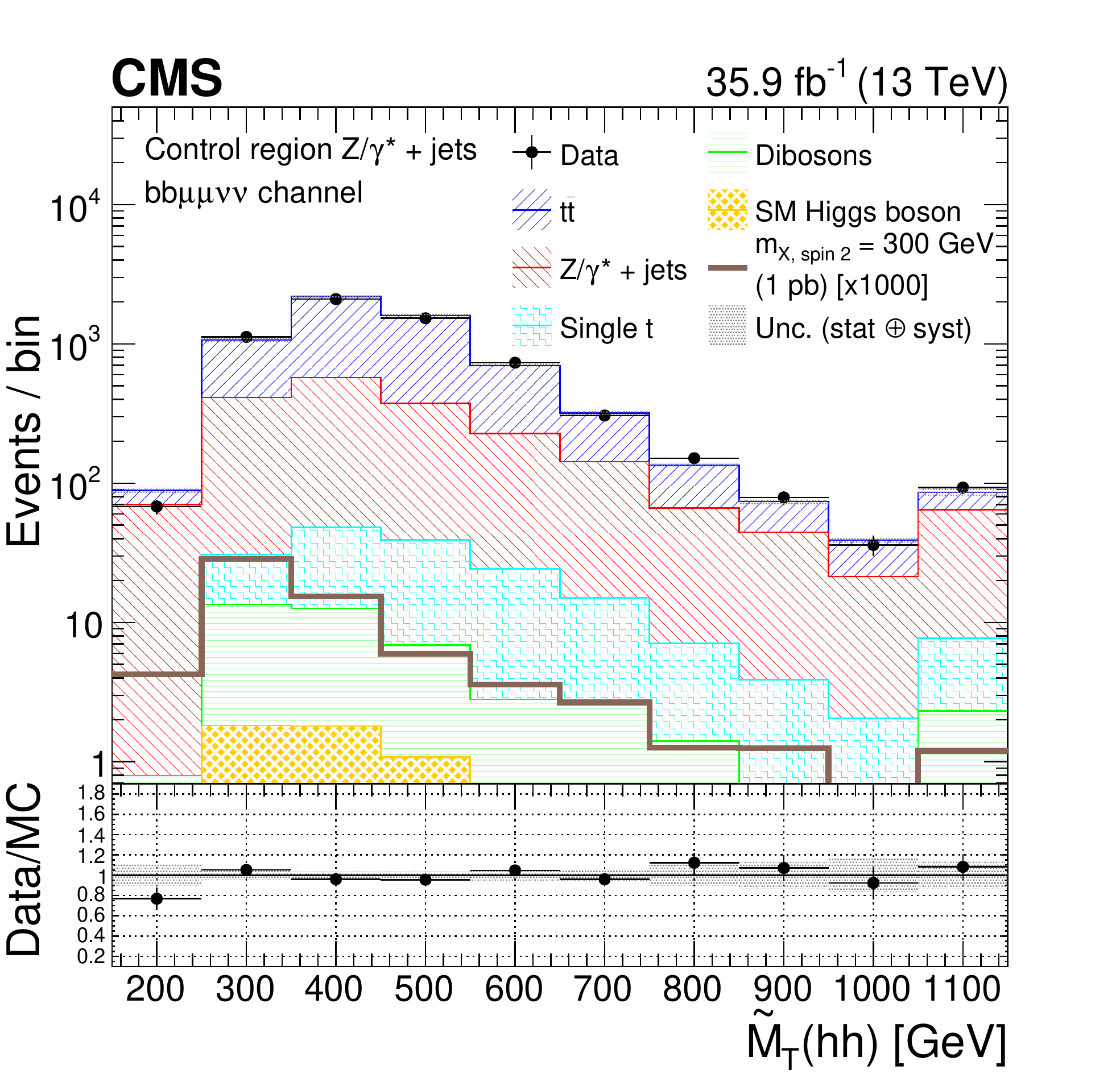}
    \includegraphics[width=0.32\textwidth]{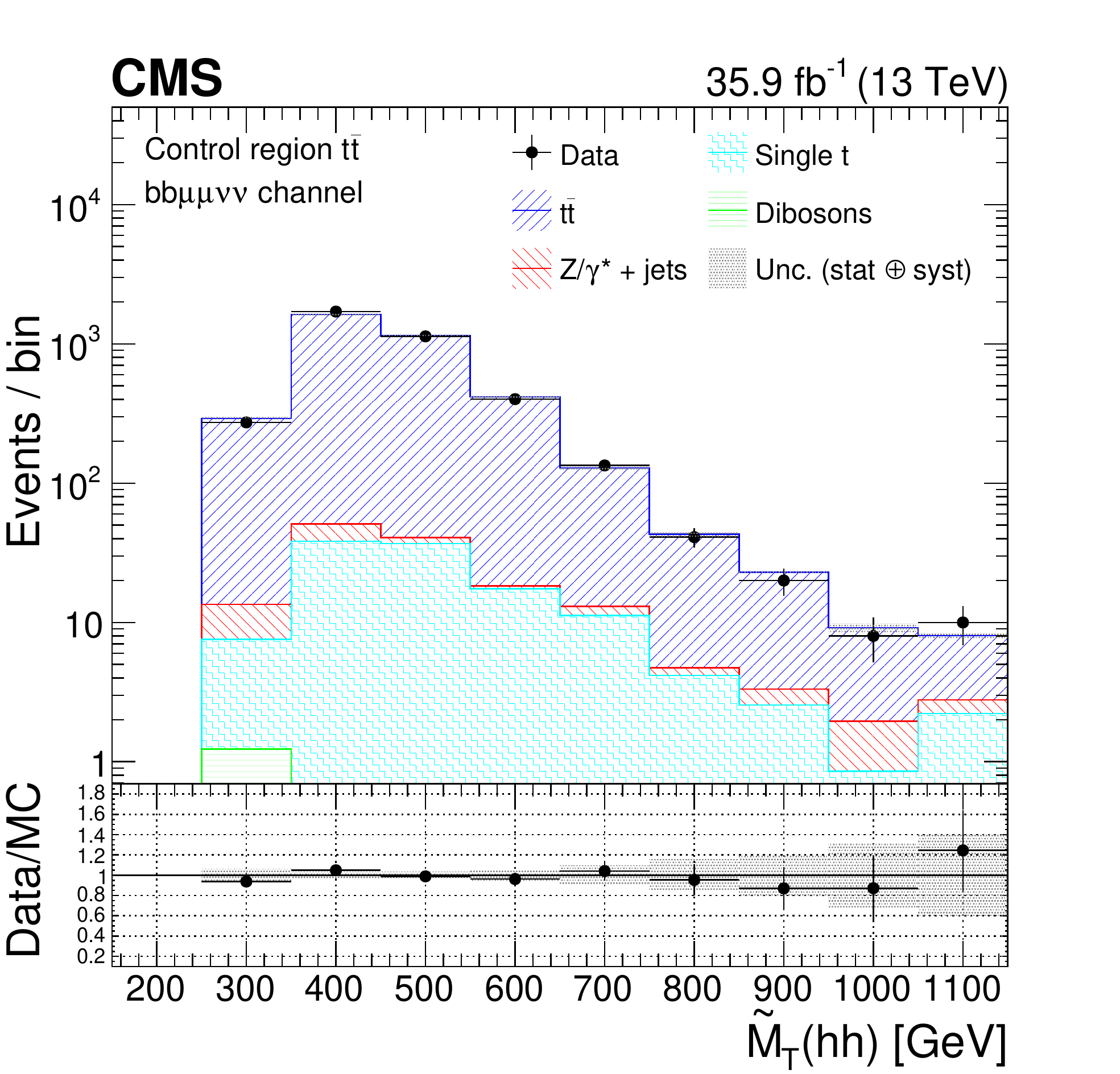}
    \includegraphics[width=0.32\textwidth]{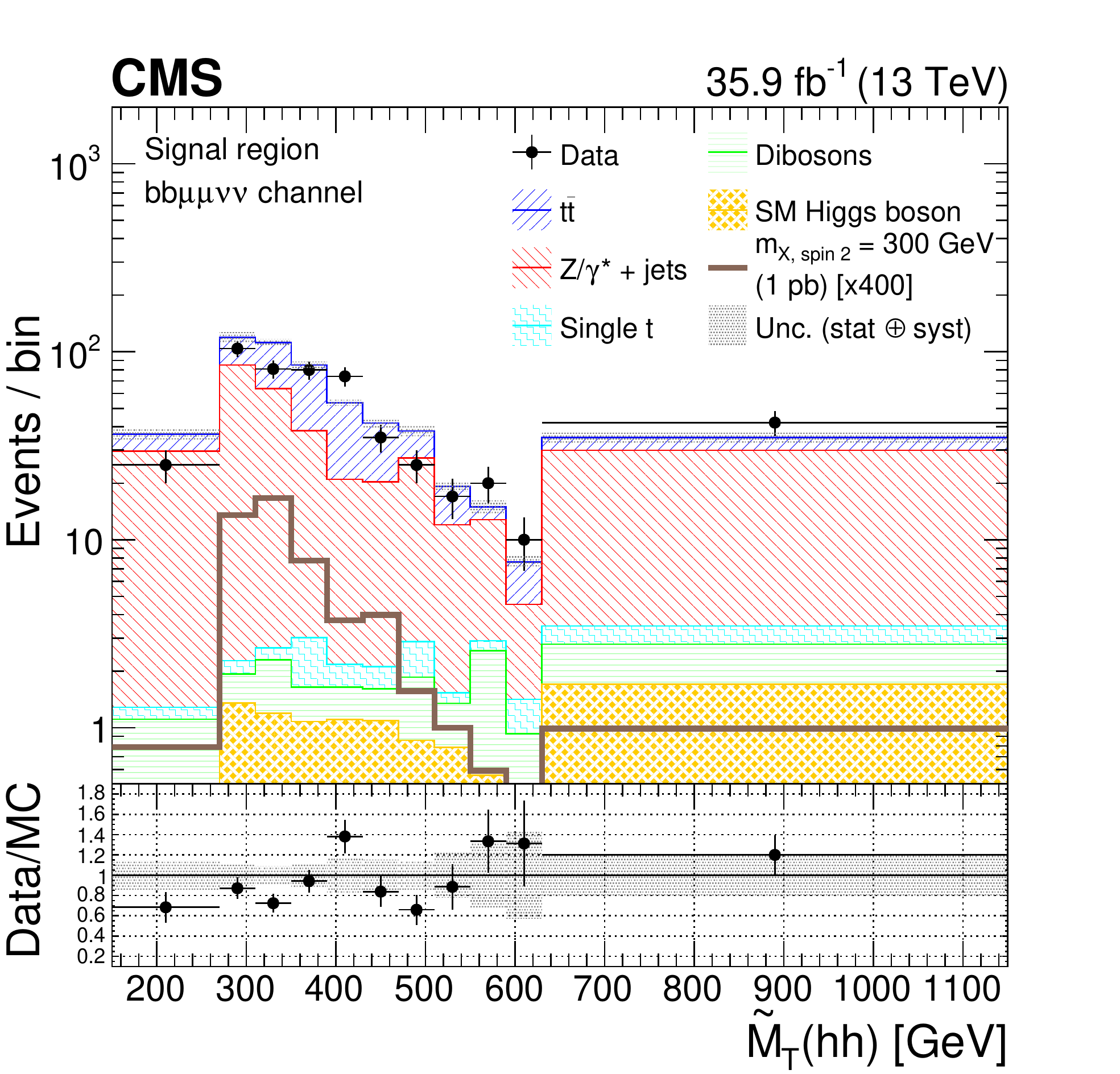} \\
    \includegraphics[width=0.32\textwidth]{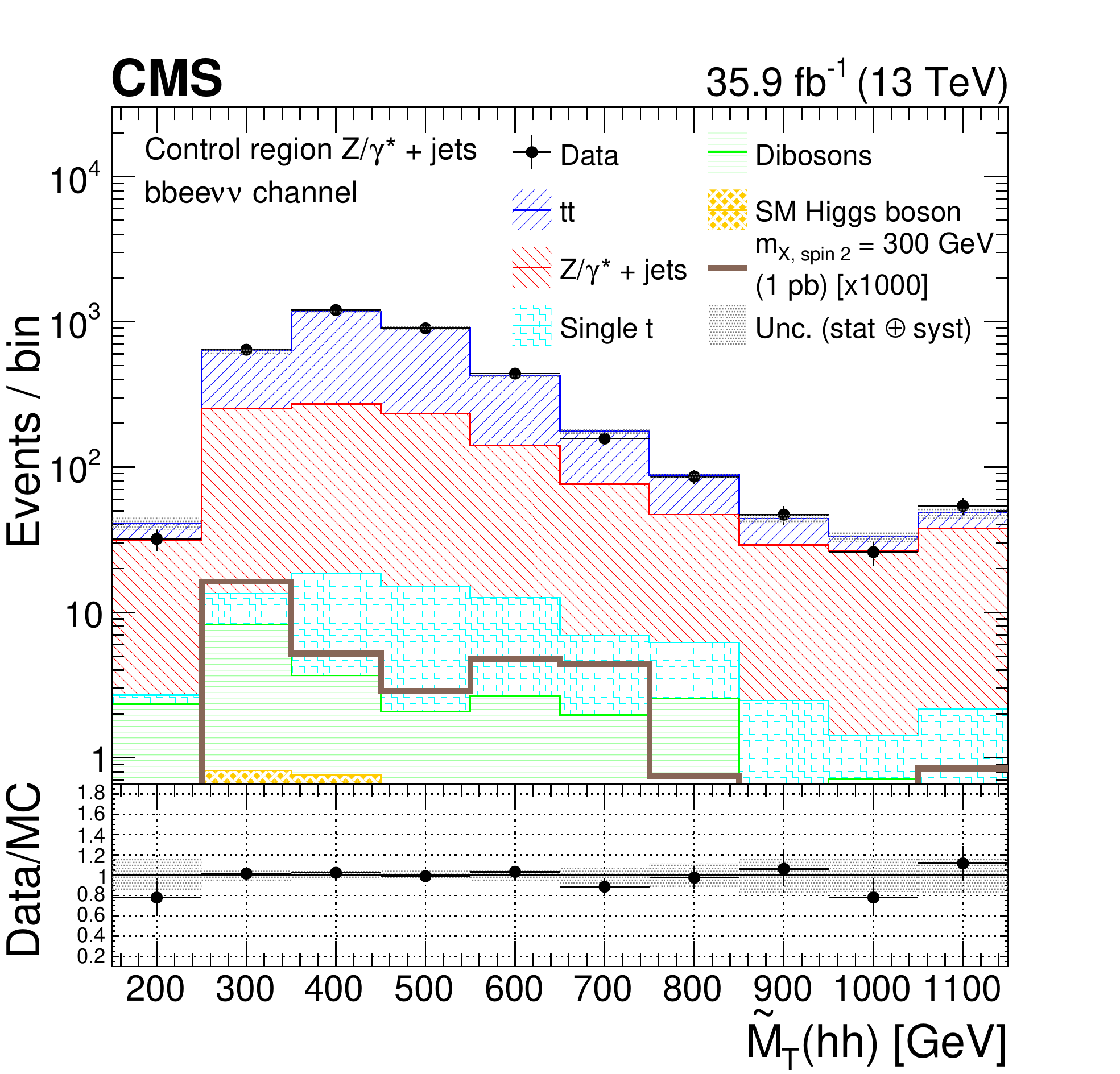}
    \includegraphics[width=0.32\textwidth]{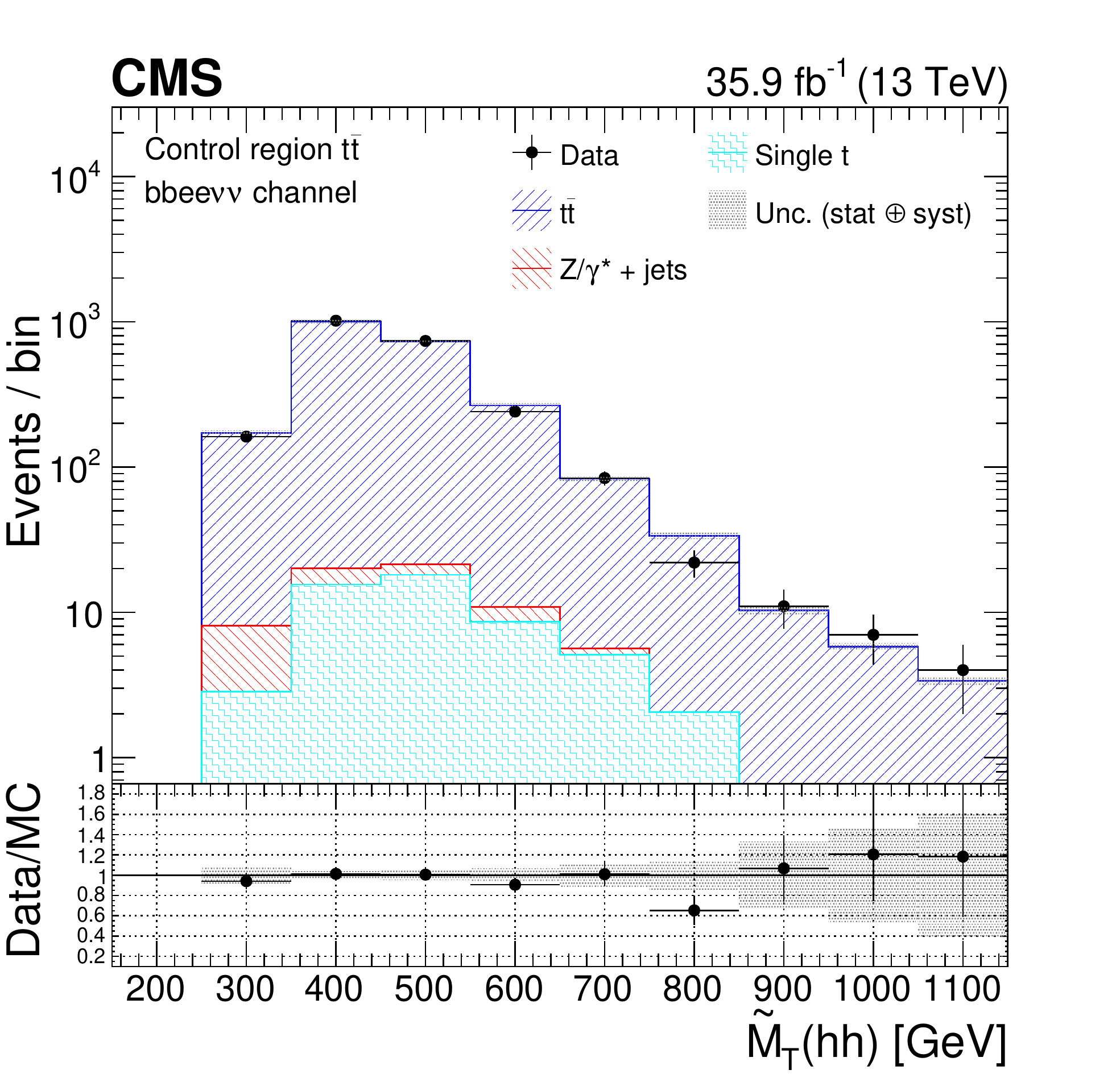}
    \includegraphics[width=0.32\textwidth]{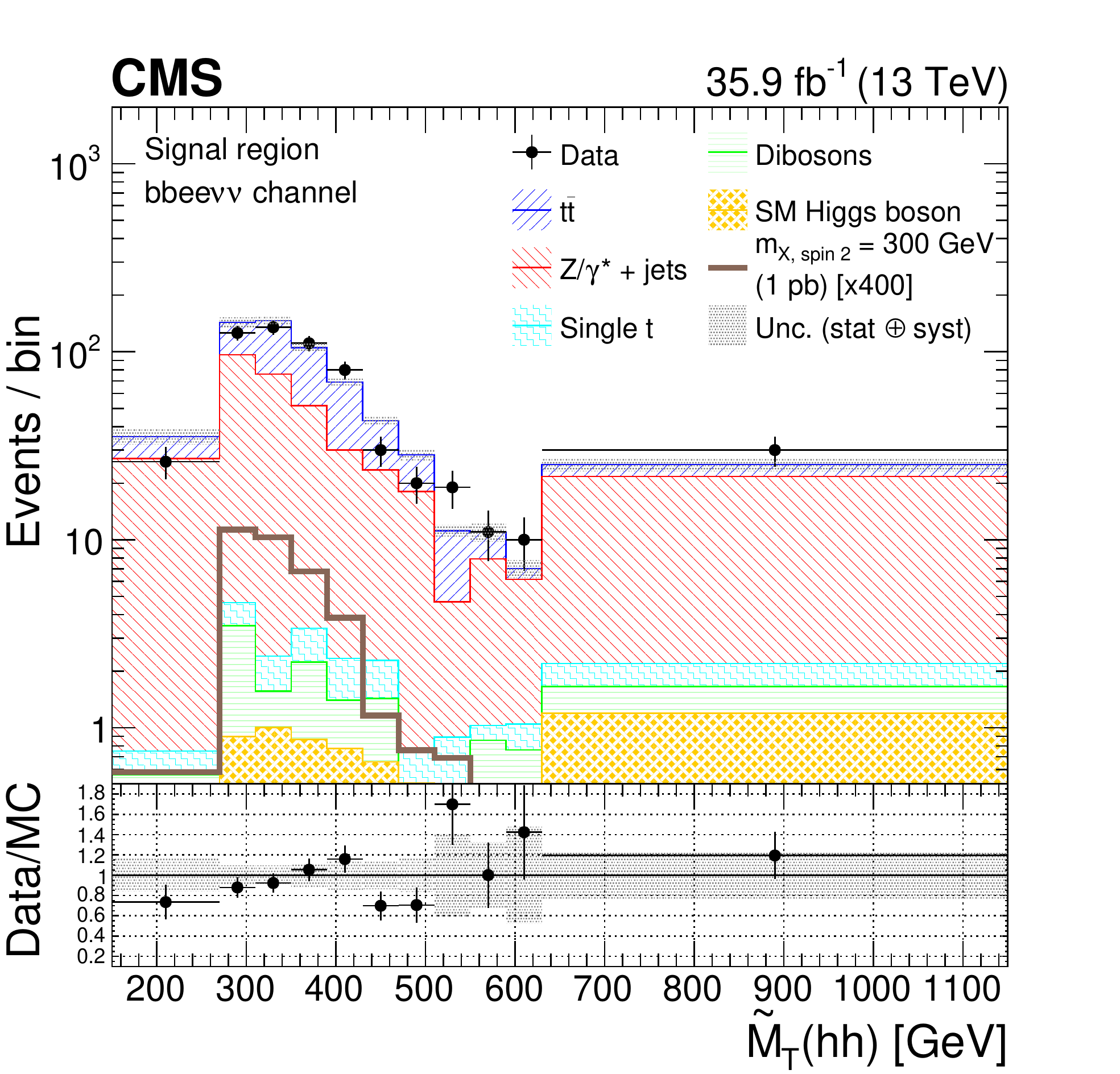}
    \caption{Pseudo transverse mass of the reconstructed $\Ph\Ph$ candidates, in the \llnunu channel, for data, simulated spin-2 RS1 graviton signal with a mass of 300\GeV, and simulated backgrounds scaled according to the fit results. The upper and lower
    rows correspond to the muon and electrons channels. For each row,
    the left and middle plots are for the \dyJets  and \ttbar control regions, and the right
    is for the signal region.  The signals are normalized to 1\unit{pb} for the $\Pp\Pp\to\mathrm{X}\to\Ph\Ph$ process. 
    The shaded area represents the combined statistical and systematic uncertainties in the background estimate.}
    \label{MCcomparisons}
\end{figure*}

Figure~\ref{MCcomparisons} shows the $\Ph\Ph$ pseudo transverse mass distributions in the data, background estimates, and spin-2 RS1 graviton for the 300\GeV mass hypothesis, after the final selection in the \llnunu channel. 

The systematic uncertainties are represented by nuisance parameters that are varied in the fit according to their probability density functions, prescribed as follows. A log-normal probability density function is assumed for the nuisance parameters affecting the event yields of the various background contributions, whereas systematic uncertainties that affect the distributions are represented by nuisance parameters whose variation is a vertical interpolation in each bin with a sixth-order polynomial for upward and downward shifts of one standard deviation, and linearly outside of that~\cite{ATLAS:2011tau}.

The statistical uncertainty from the limited number of events in the simulated samples is taken into account, for each bin of the discriminant distributions, by assigning a nuisance parameter to scale the sum of the process yields in that bin according to the statistical uncertainty using the Barlow--Beeston ``lite'' prescription~\cite{BARLOW1993219,Conway:2011in}.

In both channels the data distributions are well reproduced by the SM background processes. Upper limits on the resonance production cross section are set, using the asymptotic \CLs modified frequentist approach~\cite{Junk:1999kv,Read_2002,AsympOpt}. 

\begin{figure*}[!htbp]
       \centering
       {\includegraphics[width=.49\textwidth]{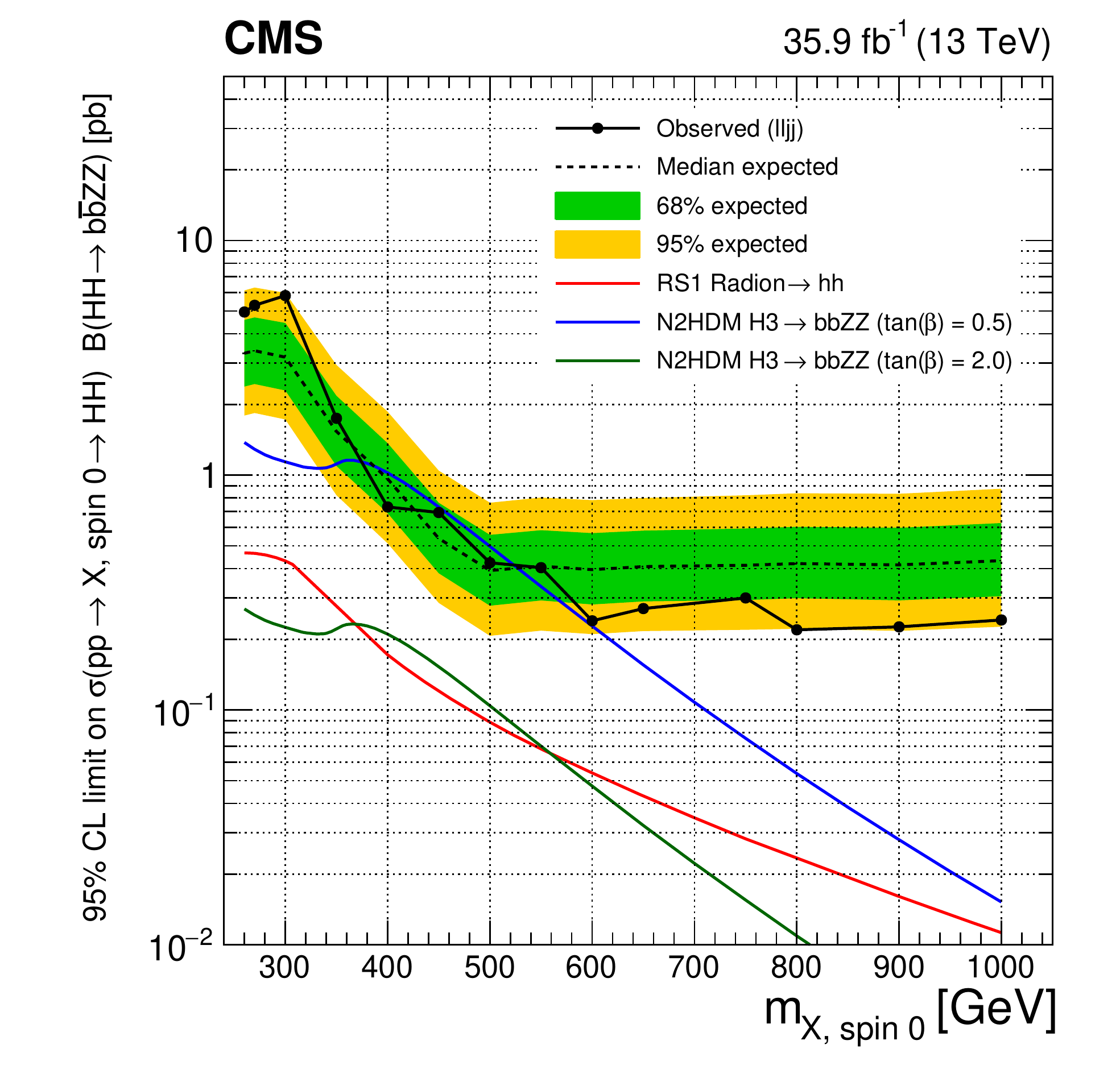}}
       {\includegraphics[width=.49\textwidth]{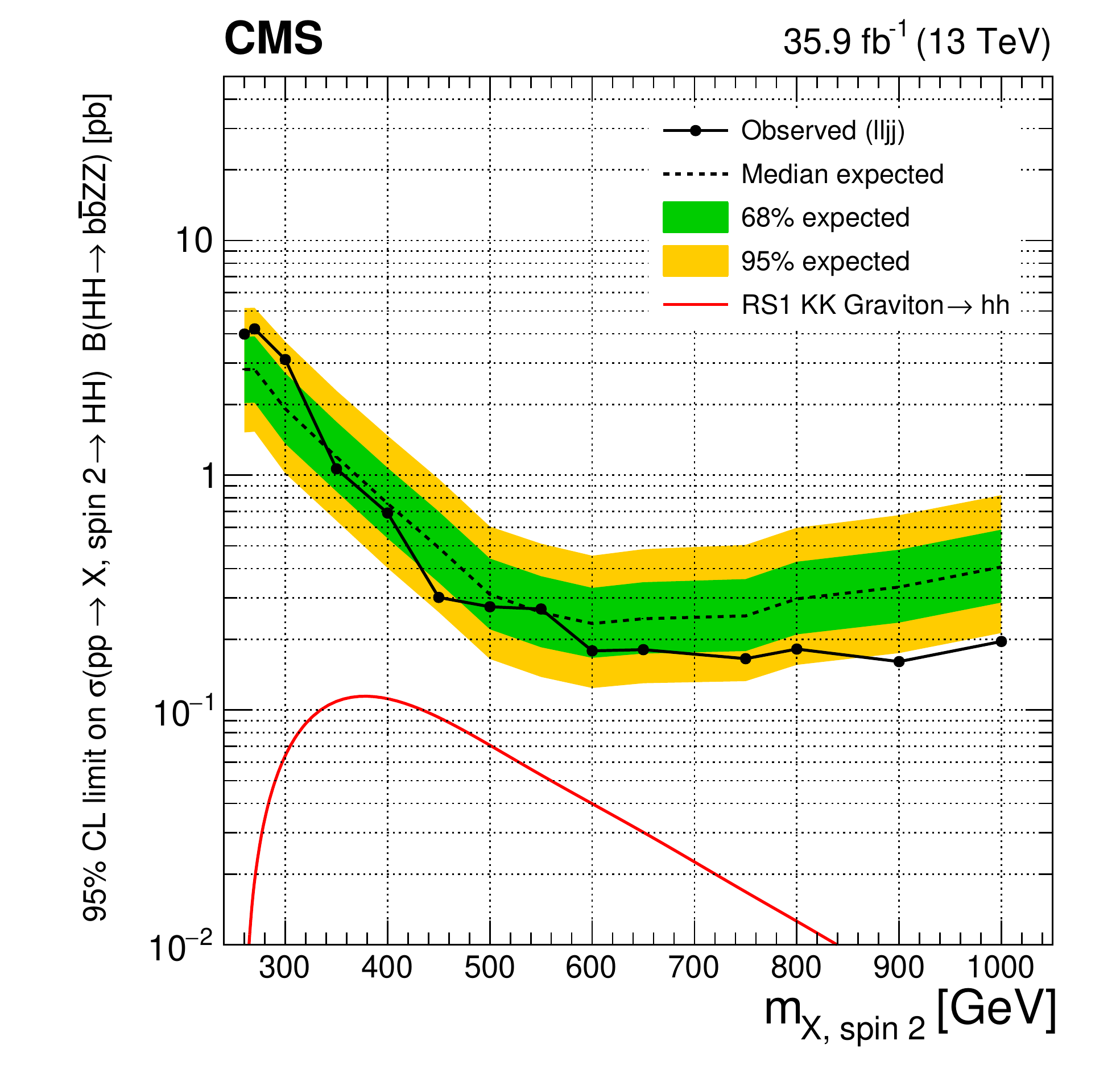}}
       {\includegraphics[width=0.49\textwidth]{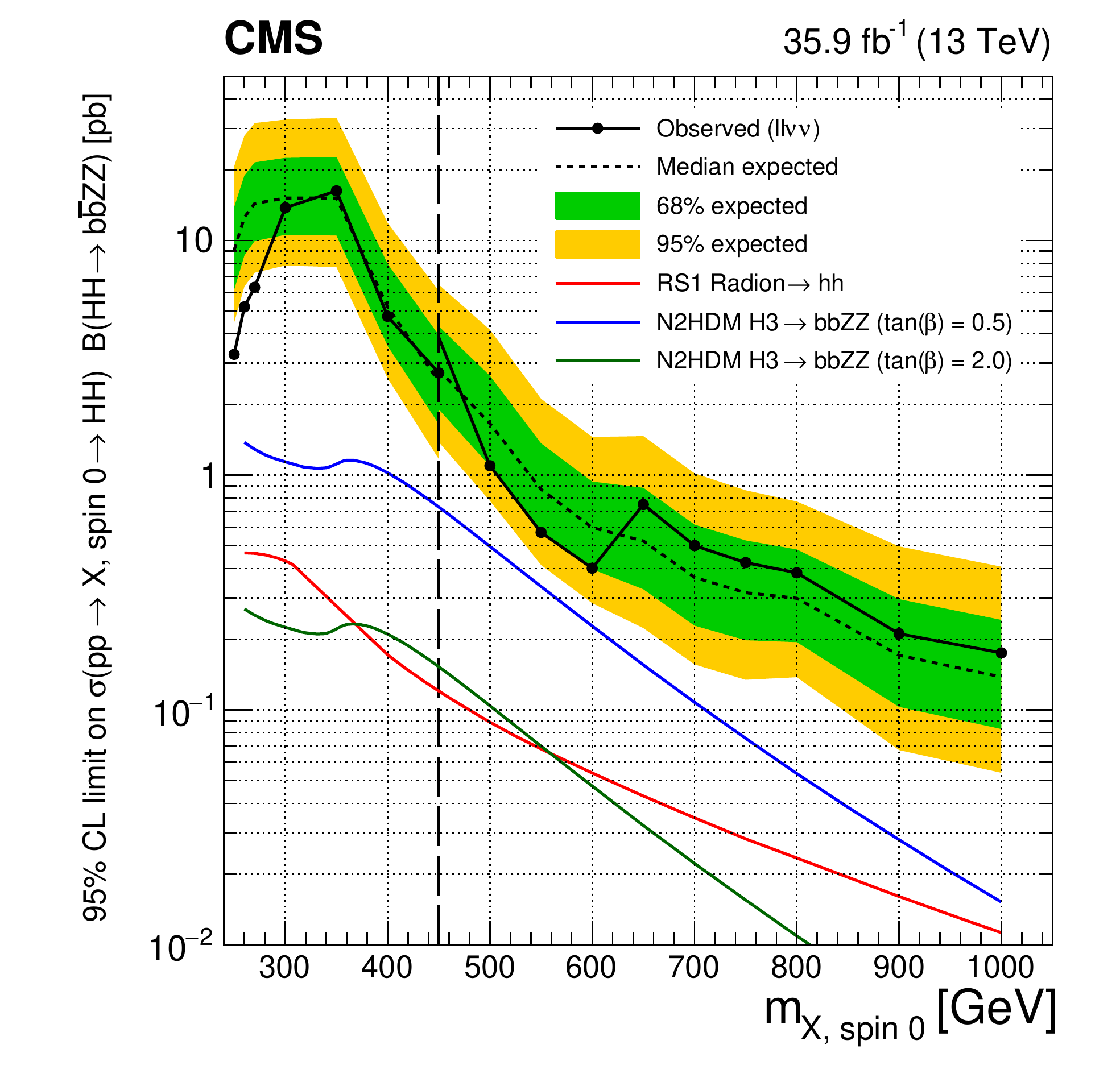}}
       {\includegraphics[width=0.49\textwidth]{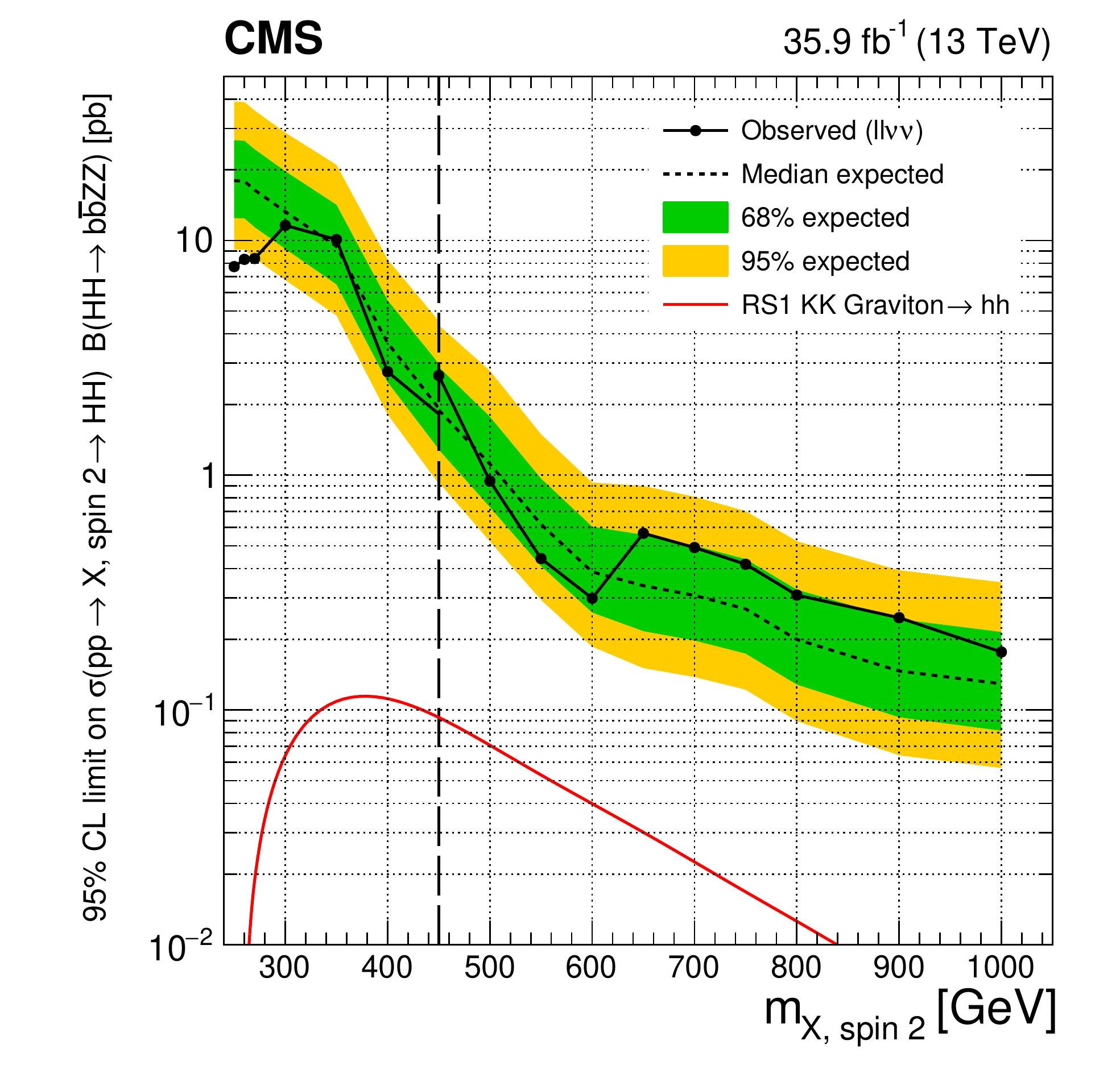}}
       \caption{
Expected (black dashed line) and observed (black solid line) limits on the cross section of resonant $\PH\PH$ production times the branching fraction of $\PH\PH\to\bbZZ$ as a function of the resonance mass for the \lljj (upper row) and \llnunu (lower row) channels, where $\PH$ can represent either the SM Higgs boson or an additional Higgs boson from an extended electroweak sector. The spin-0 case is shown on the left and the spin-2 case is shown on the right. The red solid line shows the theoretical prediction for the cross section of an RS1 radion with $\lambda_{\mathrm{R}}=1\TeV$ and $kL=35$ (left) and an RS1 KK graviton with $\tilde{k} = 0.1$ (right). In the spin-0 case only, the blue (green) line shows the decays of ${\PH}_3\to{\PH}_1{\PH}_1/{\PH}_1{\PH}_2/{\PH}_2{\PH}_2\to\bbZZ$ in the N2HDM formulation, with $\tan\beta=0.5\ (2.0)$, the scalar ${\PH}_3$ vev set to 45\GeV, and the mixing angles $\alpha_1$, $\alpha_2$, $\alpha_3$ set to 0.76, 0.48, and 1.00, respectively. The correction factor based on the relative partial width of ${\PH}_3$ to two gluons is around 3.0 (0.7) for $\tan\beta=0.5\ (2.0)$. In the lower row, the vertical black dashed line indicates the resonance mass of 450\GeV, a mass point where the BDT used in the analysis is switched from the one trained for low mass resonance to the one trained for high mass resonance.
       }
       \label{fig:limit_plots_channel}
\end{figure*}

The observed and expected 95\% \CL upper limits on $\sigma(\Pp\Pp\to\mathrm{X}\to\PH\PH\to\bbZZ)$ in the \lljj and \llnunu channels as a function of $m_X$ are shown in Fig.~\ref{fig:limit_plots_channel}, together with the NLO predictions for the RS1 radion, RS1 KK graviton, and N2HDM resonance production cross sections, where $\PH$ can represent either the SM Higgs boson or an additional Higgs boson from an extended electroweak sector. As two different BDTs are defined for the search in the low- and high-mass ranges of the \llnunu channel, the limit calculation is performed with both of the BDTs at the boundary of the two ranges, around 450\GeV, where a discontinuity is seen. 

\begin{figure*}[!htbp]
       \centering
       {\includegraphics[width=.49\textwidth]{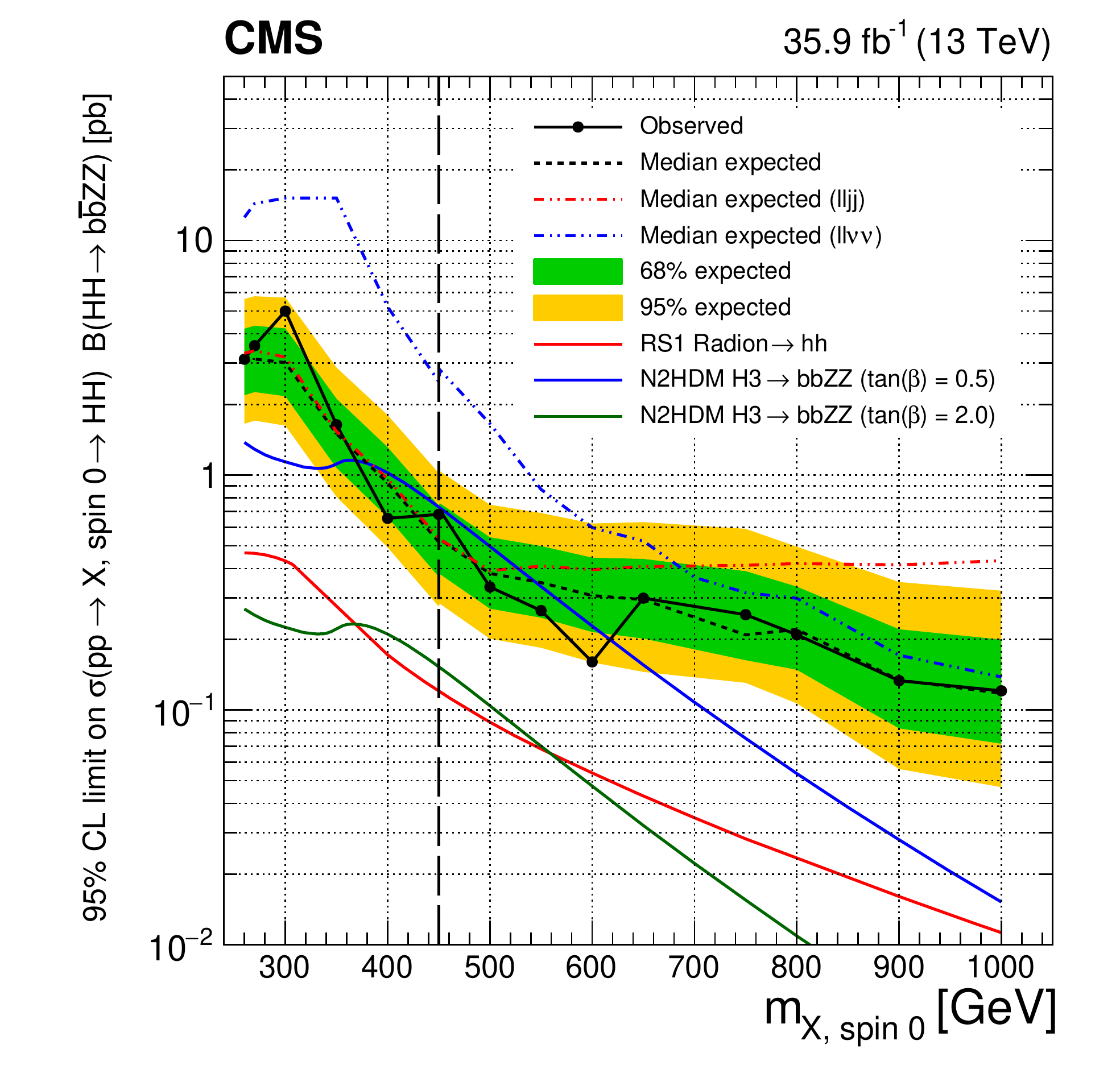}}
        {\includegraphics[width=.49\textwidth]{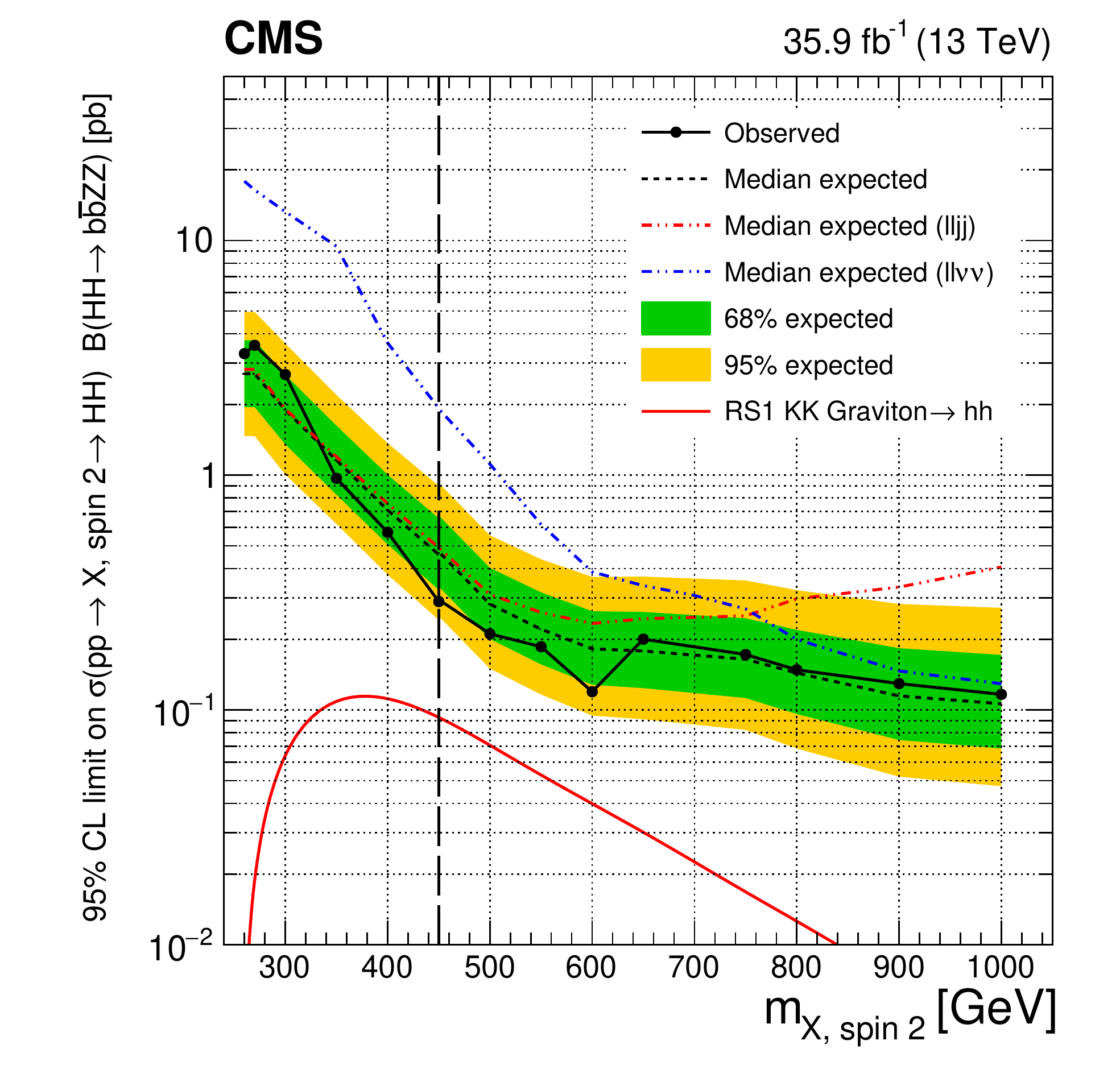}}
       \caption{
Expected (black dashed line) and observed (black solid line) limits on the cross section of resonant $\PH\PH$ production times the branching fraction of $\PH\PH\to\bbZZ$ as a function of the mass of the resonance for the combination of the \lljj and \llnunu channels, where $\PH$ can represent either the SM Higgs boson or an additional Higgs boson from an extended electroweak sector. The spin-0 case is shown on the left and the spin-2 case is shown on the right. The expected limits for each individual channel are shown with a red dashed line for the \lljj channel and blue dashed line for the \llnunu channel. The red solid lines show the theoretical prediction for the cross section of an RS1 radion with $\lambda_{\mathrm{R}}=1\TeV$ and $kL=35$ (left) and an RS1 KK graviton with $\tilde{k} = 0.1$ (right). In the spin-0 case only, the blue (green) line shows the decays of ${\PH}_3\to{\PH}_1{\PH}_1/{\PH}_1{\PH}_2/{\PH}_2{\PH}_2\to\bbZZ$ in the N2HDM formulation, with $\tan\beta=0.5\ (2.0)$, the scalar ${\PH}_3$ vev set to 45\GeV, and the mixing angles $\alpha_1$, $\alpha_2$, $\alpha_3$ set to 0.76, 0.48, and 1.00, respectively. The correction factor based on the relative partial width of ${\PH}_3$ to two gluons is around 3.0 (0.7) for $\tan\beta=0.5\ (2.0)$. The vertical black dashed line indicates the resonance mass of 450\GeV, a mass point where the BDT used in the analysis is switched from the one trained for low mass resonance to the one trained for high mass resonance.
          }
	  \label{fig:limit_plots_combined}
\end{figure*}

Combined 95\% \CL upper limits from both channels on $\sigma(\Pp\Pp\to\mathrm{X}\to\PH\PH\to\bbZZ$) as a function of $m_X$, are shown in Fig.~\ref{fig:limit_plots_combined}, together with the theoretical predictions for the RS1 radion and RS1 KK graviton. In the $m_X$ range between 260 and 1000\GeV, limits on the production cross section times branching fraction of RS1 radion and RS1 KK graviton range from 0.1 to 5.0 and 0.1 to 3.6\unit{pb}, respectively. In the spin-0 case, the predictions of the N2HDM model with $\tan\beta=0.5\mathrm{\ and\ }2.0$ are shown, for all ${\PH}_3\to{\PH}_1{\PH}_1/{\PH}_1{\PH}_2/{\PH}_2{\PH}_2\to\bbZZ$ decays. In the $\tan\beta=0.5$ case, the model can be excluded with ${\PH}_3$ in the $m_X$ range of 360--620\GeV. In comparison to previous searches in other channels, we achieve a sensitivity to the RS1 radion and RS1 KK graviton models that is consistent with the lower value of the $\Ph\Ph$ branching fraction in the \bbZZ channel relative to the other channels.

\begin{figure*}[!htbp]
       \centering
       {\includegraphics[width=.49\textwidth]{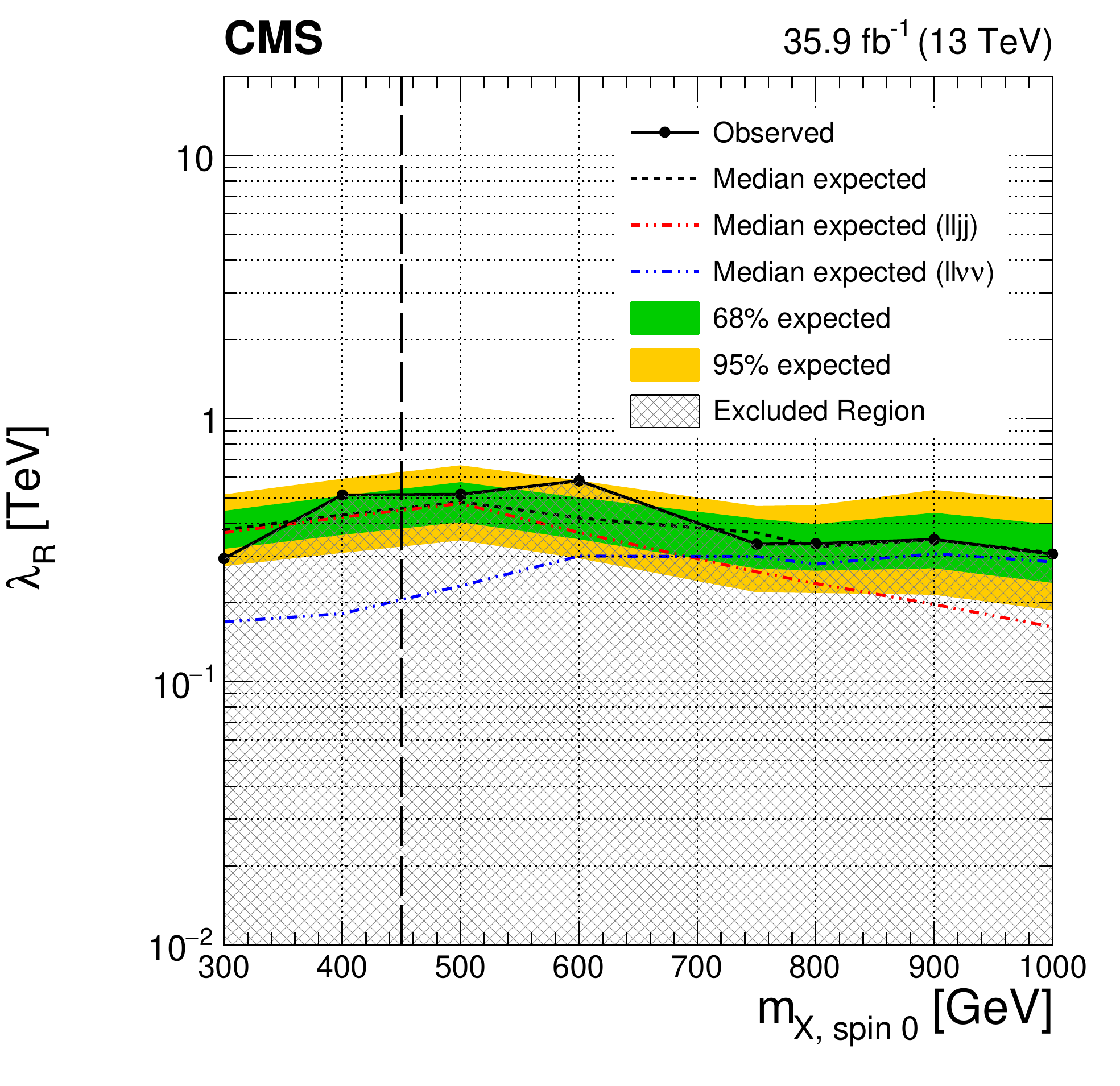}}
        {\includegraphics[width=.49\textwidth]{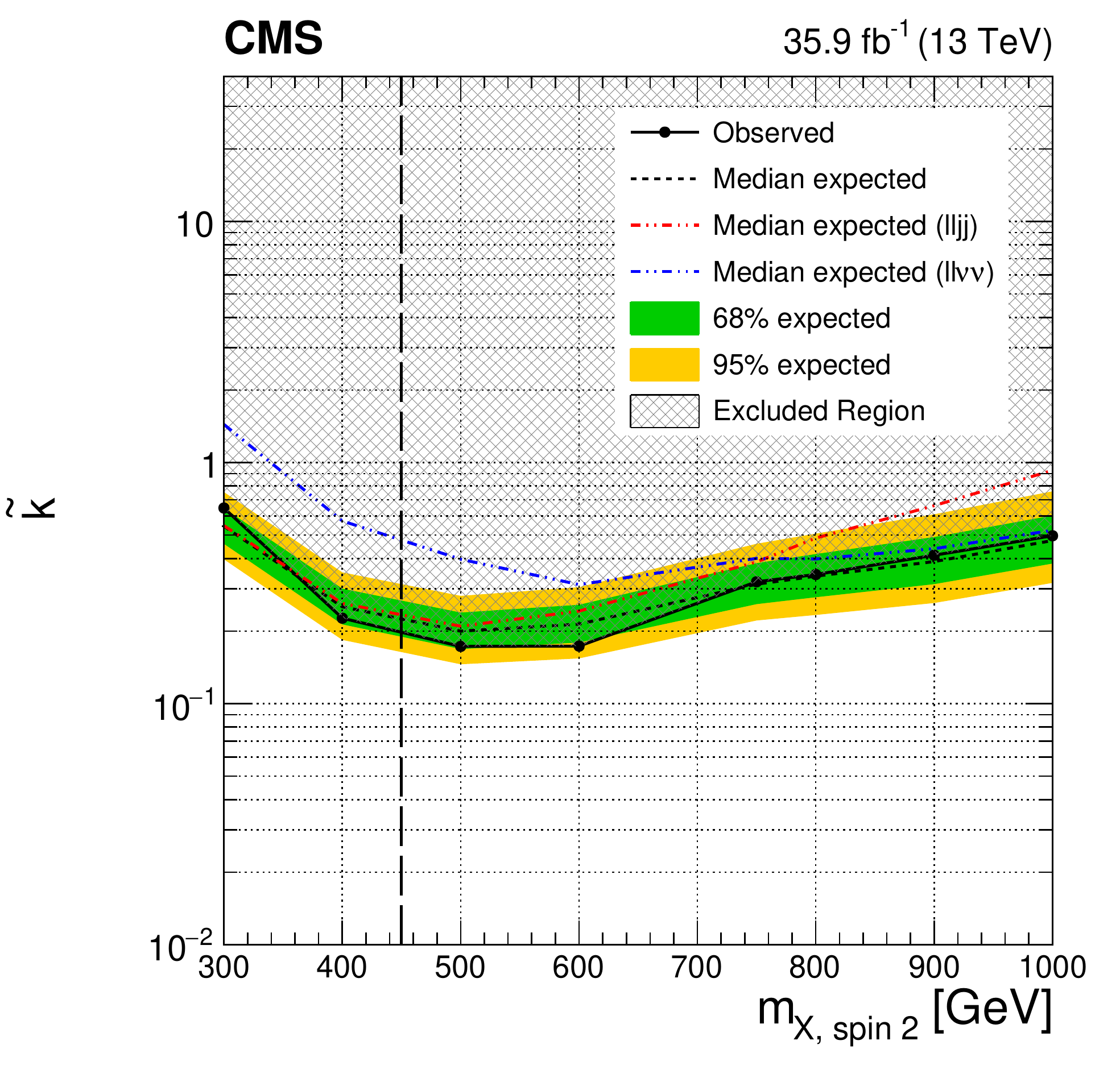}}
       \caption{
The expected and observed exclusion limits at 95\% \CL on the RS1 radion with $kL=35$ (RS1 KK graviton) hypothesis in the $\lambda_{\mathrm{R}}$ ($\tilde{k}$) versus mass plane for the individual \lljj (red) and \llnunu (blue) channels and their combination (black). The dark green and light yellow expected limit uncertainty bands represent the 68 and 95\% confidence intervals. Solid lines represent the observed limits and dashed lines represent the expected limits. The shaded region is excluded by the current limits. The vertical black dashed line indicates the resonance mass of 450\GeV, a mass point where the BDT used in the \llnunu analysis is switched from the one trained for low mass resonance to the one trained for high mass resonance.
          }
	  \label{fig:limit_plots_combined_2D}
\end{figure*}

Finally, the results are also interpreted as a function of both the $m_X$ and $\lambda_{\mathrm{R}}$ ($\tilde{k}$) for the radion (graviton) case, with $\lambda_{\mathrm{R}}<0.3\TeV$ ($\tilde{k}>0.6$) excluded for all of the $m_X$ considered, as shown in Fig.~\ref{fig:limit_plots_combined_2D}.

\section{Summary}
\label{summary}

A search for the production of a narrow-width resonance decaying into a pair of Higgs bosons decaying into the \bbZZ channel is presented. The analysis is based on data collected with the CMS detector during 2016, in proton-proton collisions at the LHC, corresponding to an integrated luminosity of 35.9\fbinv. The final states considered are the ones where one of the \PZ bosons decays into a pair of muons or electrons, and the other \PZ boson decays either to a pair of quarks or a pair of neutrinos. Upper limits at 95\% confidence level are placed on the production of narrow-width spin-0 or spin-2 particles decaying to a pair of Higgs bosons, in models with and without an extended Higgs sector.  For a resonance mass range between 260 and 1000\GeV, limits on the production cross section times branching fraction of a spin-0 and spin-2 resonance range from 0.1 to 5.0\unit{pb} and 0.1 to 3.6\unit{pb}, respectively.  These results set limits in parameter space in bulk Randall--Sundrum radion, Kaluza--Klein excitation of the graviton, and N2HDM models. For specific choices of parameters the N2HDM can be excluded in a mass range between 360 and 620\GeV for a resonance decaying to two Higgs bosons. This is the first search for Higgs boson resonant pair production in the \bbZZ  channel.

\begin{acknowledgments}
       We congratulate our colleagues in the CERN accelerator departments for the excellent performance of the LHC and thank the technical and administrative staffs at CERN and at other CMS institutes for their contributions to the success of the CMS effort. In addition, we gratefully acknowledge the computing centers and personnel of the Worldwide LHC Computing Grid for delivering so effectively the computing infrastructure essential to our analyses. Finally, we acknowledge the enduring support for the construction and operation of the LHC and the CMS detector provided by the following funding agencies: BMBWF and FWF (Austria); FNRS and FWO (Belgium); CNPq, CAPES, FAPERJ, FAPERGS, and FAPESP (Brazil); MES (Bulgaria); CERN; CAS, MoST, and NSFC (China); COLCIENCIAS (Colombia); MSES and CSF (Croatia); RPF (Cyprus); SENESCYT (Ecuador); MoER, ERC IUT, PUT and ERDF (Estonia); Academy of Finland, MEC, and HIP (Finland); CEA and CNRS/IN2P3 (France); BMBF, DFG, and HGF (Germany); GSRT (Greece); NKFIA (Hungary); DAE and DST (India); IPM (Iran); SFI (Ireland); INFN (Italy); MSIP and NRF (Republic of Korea); MES (Latvia); LAS (Lithuania); MOE and UM (Malaysia); BUAP, CINVESTAV, CONACYT, LNS, SEP, and UASLP-FAI (Mexico); MOS (Montenegro); MBIE (New Zealand); PAEC (Pakistan); MSHE and NSC (Poland); FCT (Portugal); JINR (Dubna); MON, RosAtom, RAS, RFBR, and NRC KI (Russia); MESTD (Serbia); SEIDI, CPAN, PCTI, and FEDER (Spain); MOSTR (Sri Lanka); Swiss Funding Agencies (Switzerland); MST (Taipei); ThEPCenter, IPST, STAR, and NSTDA (Thailand); TUBITAK and TAEK (Turkey); NASU (Ukraine); STFC (United Kingdom); DOE and NSF (USA). 

       \hyphenation{Rachada-pisek} Individuals have received support from the Marie-Curie program and the European Research Council and Horizon 2020 Grant, contract Nos.\ 675440, 752730, and 765710 (European Union); the Leventis Foundation; the A.P.\ Sloan Foundation; the Alexander von Humboldt Foundation; the Belgian Federal Science Policy Office; the Fonds pour la Formation \`a la Recherche dans l'Industrie et dans l'Agriculture (FRIA-Belgium); the Agentschap voor Innovatie door Wetenschap en Technologie (IWT-Belgium); the F.R.S.-FNRS and FWO (Belgium) under the ``Excellence of Science -- EOS" -- be.h project n.\ 30820817; the Beijing Municipal Science \& Technology Commission, No. Z191100007219010; the Ministry of Education, Youth and Sports (MEYS) of the Czech Republic; the Deutsche Forschungsgemeinschaft (DFG) under Germany's Excellence Strategy -- EXC 2121 ``Quantum Universe" -- 390833306; the Lend\"ulet (``Momentum") Program and the J\'anos Bolyai Research Scholarship of the Hungarian Academy of Sciences, the New National Excellence Program \'UNKP, the NKFIA research grants 123842, 123959, 124845, 124850, 125105, 128713, 128786, and 129058 (Hungary); the Council of Science and Industrial Research, India; the HOMING PLUS program of the Foundation for Polish Science, cofinanced from European Union, Regional Development Fund, the Mobility Plus program of the Ministry of Science and Higher Education, the National Science Center (Poland), contracts Harmonia 2014/14/M/ST2/00428, Opus 2014/13/B/ST2/02543, 2014/15/B/ST2/03998, and 2015/19/B/ST2/02861, Sonata-bis 2012/07/E/ST2/01406; the National Priorities Research Program by Qatar National Research Fund; the Ministry of Science and Higher Education, project no. 02.a03.21.0005 (Russia); the Tomsk Polytechnic University Competitiveness Enhancement Program and ``Nauka" Project FSWW-2020-0008 (Russia); the Programa Estatal de Fomento de la Investigaci{\'o}n Cient{\'i}fica y T{\'e}cnica de Excelencia Mar\'{\i}a de Maeztu, grant MDM-2015-0509 and the Programa Severo Ochoa del Principado de Asturias; the Thalis and Aristeia programs cofinanced by EU-ESF and the Greek NSRF; the Rachadapisek Sompot Fund for Postdoctoral Fellowship, Chulalongkorn University and the Chulalongkorn Academic into Its 2nd Century Project Advancement Project (Thailand); the Kavli Foundation; the Nvidia Corporation; the SuperMicro Corporation; the Welch Foundation, contract C-1845; and the Weston Havens Foundation (USA). 
\end{acknowledgments}

\bibliography{auto_generated}  

\cleardoublepage \appendix\section{The CMS Collaboration \label{app:collab}}\begin{sloppypar}\hyphenpenalty=5000\widowpenalty=500\clubpenalty=5000\vskip\cmsinstskip
\textbf{Yerevan Physics Institute, Yerevan, Armenia}\\*[0pt]
A.M.~Sirunyan$^{\textrm{\dag}}$, A.~Tumasyan
\vskip\cmsinstskip
\textbf{Institut f\"{u}r Hochenergiephysik, Wien, Austria}\\*[0pt]
W.~Adam, F.~Ambrogi, T.~Bergauer, M.~Dragicevic, J.~Er\"{o}, A.~Escalante~Del~Valle, R.~Fr\"{u}hwirth\cmsAuthorMark{1}, M.~Jeitler\cmsAuthorMark{1}, N.~Krammer, L.~Lechner, D.~Liko, T.~Madlener, I.~Mikulec, N.~Rad, J.~Schieck\cmsAuthorMark{1}, R.~Sch\"{o}fbeck, M.~Spanring, S.~Templ, W.~Waltenberger, C.-E.~Wulz\cmsAuthorMark{1}, M.~Zarucki
% ...
\vskip\cmsinstskip
\dag: Deceased\\
1:  Also at Vienna University of Technology, Vienna, Austria\\
% ...
\end{sloppypar}
\end{document}